\documentclass[twocolumn,twocolappendix]{aastex631}

\usepackage{amsmath}
\usepackage{bm}
\usepackage{microtype}
\usepackage{physics}

\newcommand{\cs}{c_{\rm s}}

\newcommand{\vinf}{v_\infty}
\newcommand{\msun}{M_\odot}
\newcommand{\rsun}{R_\odot}

\newcommand{\red}[1]{{#1}}

\defcitealias{ShiHuXu_2026arXiv260827596S}{Paper~I}

\begin{document}

\title{Dynamical formation of high-eccentricity compact binaries through BH--BH*/TZO collisions}

\correspondingauthor{Qingru Hu; Yanlong Shi}
\email{Email: huqr24@mails.tsinghua.edu.cn; yanlong@cita.utoronto.ca}

\newcommand{\cita}{Canadian Institute for Theoretical Astrophysics, University of Toronto, Toronto, ON M5S 3H8, Canada}
\newcommand{\caltech}{TAPIR, MC 350-17, California Institute of Technology, Pasadena, CA 91125, USA}

\newcommand{\ucsc}{Department of Astronomy and Astrophysics, University of California, Santa Cruz, CA, 95064, USA}

\newcommand{\tsinghuaias}{Institute for Advanced Study, Tsinghua University, Beijing, 100084, China}

\newcommand{\tsinghuadoa}{Department of Astronomy, Tsinghua University, Beijing, 100084, China}

\newcommand{\westlake}{Department of Astronomy, Westlake University, Hangzhou, Zhejiang, 310030, China}

\newcommand{\ucsb}{Department of Physics, University of California, Santa Barbara, CA 93106, USA}

\author[0009-0007-9015-9451]{Qingru Hu}
\affiliation{\tsinghuadoa}

\author[0000-0002-0087-3237]{Yanlong Shi}
\affiliation{\cita}

\author[0000-0001-5466-4628]{Douglas N. C. Lin}
\affiliation{\westlake}
\affiliation{\ucsc}
\affiliation{\tsinghuaias}

\author[0000-0002-8659-3729]{Norman Murray}
\affiliation{\cita}





\begin{abstract}

The rapidly accumulating discoveries of binary stellar-mass black-hole (sBH) coalescences, detected by LIGO, have opened a new window into the formation and evolution of compact binaries. In particular, residual orbital eccentricity may provide a distinctive signature of their formation channels. Here, we investigate a scenario in which high-eccentricity compact binaries form through the sequential capture of multiple sBHs by massive main-sequence stars, using a combination of hydrodynamical and semianalytic few-body simulations. We find that sBHs with $M_\bullet\lesssim 0.2\,M_{\star}$ can be captured by massive stars and settle into a quasi-hydrostatic black-hole star (BH*) through gas dynamical friction. A subsequent encounter with a second sBH can then produce a compact binary embedded within the stellar envelope. Our hydrodynamical simulations show that through captures with small impact parameters, some binaries are born with high eccentricity ($e\gtrsim 0.5$), with their orbital frequencies already in the LISA band. Our semianalytic models further demonstrate that gas dynamical friction can pump the eccentricity to $e_{\rm 10\,Hz}>0.9$ before gravitational-wave emission eventually circularizes the binary during the final stage of coalescence. Once formed, the binary can merge quickly in $\sim 10$ hours. This channel may operate in dense stellar environments, such as star clusters and active galactic nucleus (AGN) disks. The same mechanism can also be applied to Thorne-\.Zytkow objects.  A high-eccentricity binary in the LIGO band could therefore provide a distinctive signature of this formation scenario.

\end{abstract}



\section{Introduction} 
\label{sec:intro}

Since the first detection of a binary black hole (BBH) merger in 2015 \citep[][]{AbbottAbbottAbbott_2016PhRvL.116f1102A}, the LIGO/Virgo/KAGRA (LVK) collaboration has cataloged hundreds of confirmed gravitational-wave (GW) events from compact binary mergers, elevating the population of merging compact objects from theoretical prediction to an observational reality.
The standard analyses underlying these catalogs model the binaries as quasi-circular: the waveform templates assume zero orbital eccentricity, and eccentricity is not included in the parameter-estimation products \citep[e.g.,][]{TheLIGOScientificCollaborationtheVirgoCollaborationtheKAGRACollaboration_2026arXiv260527226T}. This assumption is well motivated for most sources, because GW emission efficiently circularizes binary orbits: a binary formed with moderate eccentricity will have its orbit substantially circularized long before $f_{\rm GW}$ enters the sensitive band ($\sim 10\,\rm Hz$) of LVK detectors \citep{Peters_1964PhRv..136.1224P,SamsingLeighTrani_2018MNRAS.481.5436S}. Nonetheless, nonzero eccentricity has been claimed for several individual events \citep[][]{Romero-ShawLaskyThrane_2021ApJ...921L..31R,Romero-ShawLaskyThrane_2022ApJ...940..171R,GupteRamos-BuadesBuonanno_2025PhRvD.112j4045G}: notable examples are GW190521 \citep[mass-gap BHs;][]{Romero-ShawLaskyThrane_2020ApJ...903L...5R,GayathriHealyLange_2022NatAs...6..344G,GambaBreschiCarullo_2023NatAs...7...11G} and GW200105 \citep[NSBH;][]{PlanasHusaRamos-Buades_2025ApJ...995...47P,MorrasPrattenSchmidt_2026ApJ..1000L...2M,PhukonSchmidtMorras_2026PhRvD.113j3023P,JanTsaoO'Shaughnessy_2026PhRvD.113b4018J}.

Orbital eccentricity is nonetheless one of the most robust tracers of a compact binary's formation channel. Because GW emission circularizes orbits so efficiently, a binary that retains measurable eccentricity in the $\sim 10\,\rm Hz$ band must have been driven to small separations only shortly before merger, through pathways like binary-single interactions \citep[][]{SamsingRamirez-Ruiz_2017ApJ...840L..14S}. Binaries assembled dynamically in dense environments such as star clusters \citep[][]{PortegiesZwartMcMillan_2000ApJ...528L..17P,SamsingD'Orazio_2018MNRAS.481.5445S,Samsing_2018PhRvD..97j3014S} and AGN disks \citep[][]{YangBartosGayathri_2019PhRvL.123r1101Y,TagawaKocsisHaiman_2021ApJ...907L..20T,SamsingBartosD'Orazio_2022Natur.603..237S,TagawaHaimanKocsis_2026arXiv260425994T} can meet this condition \citep[yet field triples can also form high-eccentricity binaries through Lidov-Kozai cycles, e.g.,][]{AntoniniToonenHamers_2017ApJ...841...77A}. The detection of eccentric mergers therefore implies that at least a fraction of the observed GW events originate in dynamical environments, particularly dense star clusters \citep[][]{ZevinRomero-ShawKremer_2021ApJ...921L..43Z,Romero-ShawLaskyThrane_2022ApJ...940..171R}. Identifying the specific mechanisms that can produce and preserve high eccentricity is thus central to interpreting these events.

In the dense stellar environments most commonly invoked for BBH formation, interactions are not limited to BH--BH encounters: collisions between stellar-mass BHs (sBHs) and massive stars are also expected to occur at significant rates in dense star clusters \citep[e.g.,][]{RoseNaozSari_2022ApJ...929L..22R,KirogluKremerRasio_2025ApJ...994L..37K,Rantala_2026arXiv260422924R} and AGN disks \citep[e.g.,][]{ChenLin_2024ApJ...967...88C}. In this paper, we consider the possibility that such star--sBH collisions open formation channels qualitatively distinct from purely gravitational BH--BH scattering and from the secular evolution of hierarchical systems, and that they may give rise to the high eccentricities detectable with GW observatories.

\begin{figure*}
    \centering
    \includegraphics[width=\linewidth]{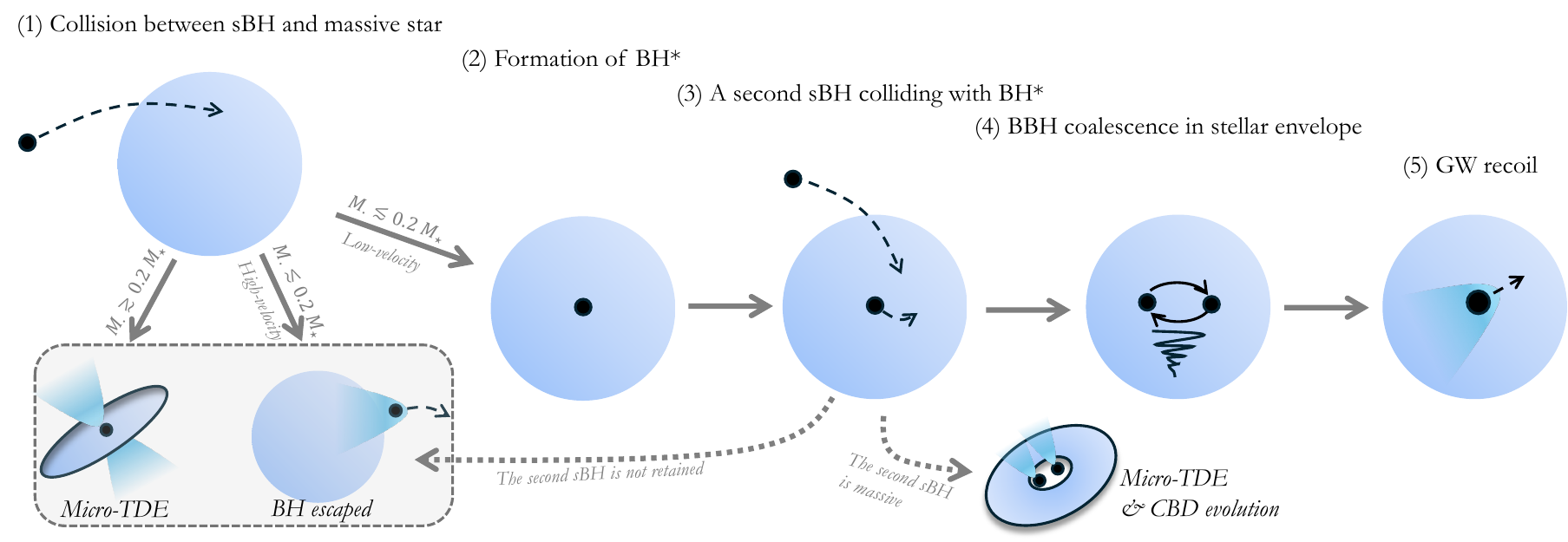}
    \caption{Illustrative summary of the scenario to form compact stellar-mass BBHs, which may happen in dense star clusters and AGN disks. The scenario initializes with the formation of a BH* in hydro-equilibrium, which requires a low-mass ($M_\bullet\lesssim 0.2\,M_\star$) sBH colliding with the massive star at low velocity \citep[Paper~I:][]{ShiHuXu_2026arXiv260827596S}. A similar sequential sBH colliding with the BH* can lead to the BBH formation and coalescence.  }
    \label{fig:scenario}
\end{figure*}

We illustrate this scenario in Figure~\ref{fig:scenario}. Such a channel proceeds through the formation of a ``BH star'' (BH*): a massive star with an sBH embedded at its center \citep[][]{BegelmanRossiArmitage_2008MNRAS.387.1649B,VolonteriBegelman_2010MNRAS.409.1022V,BallToutZytkow_2011MNRAS.414.2751B,CoughlinBegelman_2024ApJ...970..158C,HassanPernaCantiello_2026ApJ...998...65H}, which is conceptually similar to the Thorne-\.Zytkow object \citep[T\.ZO;][]{ThorneZytkow_1977ApJ...212..832T}, i.e., a neutron star engulfed by a red supergiant. In a companion paper \citep[Paper~I;][]{ShiHuXu_2026arXiv260827596S}, we studied the first step of this process: the formation of a BH* via collision between an sBH and a massive main-sequence star. Using both semianalytic and hydrodynamical simulations, we showed that low-mass sBHs ($M_\bullet \lesssim 0.2\,M_\star$) can be efficiently retained by gas drag to form a quasi-hydrostatic BH* that keeps most of the original stellar mass, if they collide with the star at a low velocity ($v_\infty\lesssim \sqrt{G M_{\star}/R_\star}$). With \texttt{MESA} modeling \citepalias[see][]{ShiHuXu_2026arXiv260827596S}, we also found that the gas-drag dissipative heating cannot trigger runaway nuclear burning. On the other hand, light sBHs may escape from the stellar envelope with a higher relative impact velocity, and more massive ($M_\bullet \gtrsim 0.2\,M_\star$) sBHs bear a high gravitational potential energy that may disrupt the star instead, leading to ``micro'' tidal disruption events \citep[micro-TDE;][]{PeretsLiLombardi_2016ApJ...823..113P,KremerLombardiLu_2022ApJ...933..203K}.

If a second sBH subsequently collides with a BH* in hydro-equilibrium, like the previous star--sBH collision, it can lose energy to gas dynamical friction drag and sink toward the center of the potential well, forming a BBH. The dense stellar envelope then provides a natural hardening mechanism through gas drag, while the impulsive collision (rather than a slow inspiral) can imprint high eccentricity on the nascent binary. The binary then evolves dominantly under the gas dynamical friction before its orbital decay becomes GW-radiation-dominated. The resulting GW signature of such a binary born and hardening inside a massive star may be qualitatively distinct from those produced by other formation channels.

In this paper we study the collision between a BH* and a second sBH, and the subsequent formation and evolution of a BBH inside the stellar envelope. We take as our primary the quasi-hydrostatic BH* characterized in \citetalias{ShiHuXu_2026arXiv260827596S}, and scan secondary sBH masses in the same low-mass range ($M_\bullet \lesssim 0.1\,M_\star$) for which \citetalias{ShiHuXu_2026arXiv260827596S} found that a bound, mass-retaining configuration forms, so that the two studies sample a consistent region of parameter space. As in \citetalias{ShiHuXu_2026arXiv260827596S}, we combine three-dimensional hydrodynamical simulations with semianalytic few-body simulations, here spanning a range of secondary BH masses, impact parameters, and initial velocities.

The remainder of this paper is organized as follows. Section~\ref{sec:methods} describes the simulation setup and initial conditions for both the hydrodynamical and semianalytic methods, together with the strengths and weaknesses of each. Section~\ref{sec:hydro} presents the hydrodynamical results, including binary formation, orbital evolution, and GW implications. Section~\ref{sec:semianalytic} presents the semianalytic treatment of the BBH evolution, with particular attention to post-Newtonian effects in the late-stage coalescence. Section~\ref{sec:discussion} discusses the astrophysical implications, and Section~\ref{sec:conclusions} summarizes our conclusions.

\section{Methods}
\label{sec:methods}

We use both hydrodynamical simulations and semianalytic few-body simulations to model the dynamical formation and further evolution of the BBH after the BH--BH* collision (cf. Figure~\ref{fig:scenario}). 

Both kinds of simulations share similar initial condition configurations, each of which includes a hydrostatic BH* and an incoming sBH. The mass of the gaseous envelope for the BH*s is fixed at $M_\star=100\,M_\odot$. As in \citetalias{ShiHuXu_2026arXiv260827596S}, the sBH and the BH* are separated by $d=2\,R_\star$, where $R_{\rm \star} = (M_\star/M_\odot)^{0.6}\,R_\odot$, roughly corresponding to the radius-mass relation of a massive main-sequence star \citep[][]{ToutPolsEggleton_1996MNRAS.281..257T}. The BH* system (made of the gaseous envelope and an enclosed ``central'' BH of mass $m_1$, so the total mass is $M_{\rm BH\star} = M_\star+m_1$) is at rest initially, while the secondary BH (of mass $m_2$) collides with the BH* at an impact velocity $\bm{v}$. The magnitude of this particular impact velocity is related to the impact velocity at infinity ($v_\infty$) through $v = [2 G(M_{\rm BH\star}+m_2)/d + v_\infty^2]^{1/2} = (v_0^2+v_\infty^2)^{1/2}$, where $v_0^2 \equiv 2 G(M_{\rm BH\star}+m_2)/d$. The direction of $\bm{v}$ is parameterized with an ``impact parameter'' $b$ through $b\equiv |\hat{\bm{v}}\times \bm{d}|$, which is further associated with the impact parameter at infinity ($b_\infty$) through $l = b (v_0^2 + v_\infty^2)^{1/2} = b_\infty v_\infty$. Since  $b_\infty \to \infty$ for $l\neq 0$ parabolic orbits, we parameterize the impact velocity with $v_\infty$ and~$b$.

\begin{figure*}
    \centering
    \includegraphics[width=\textwidth]{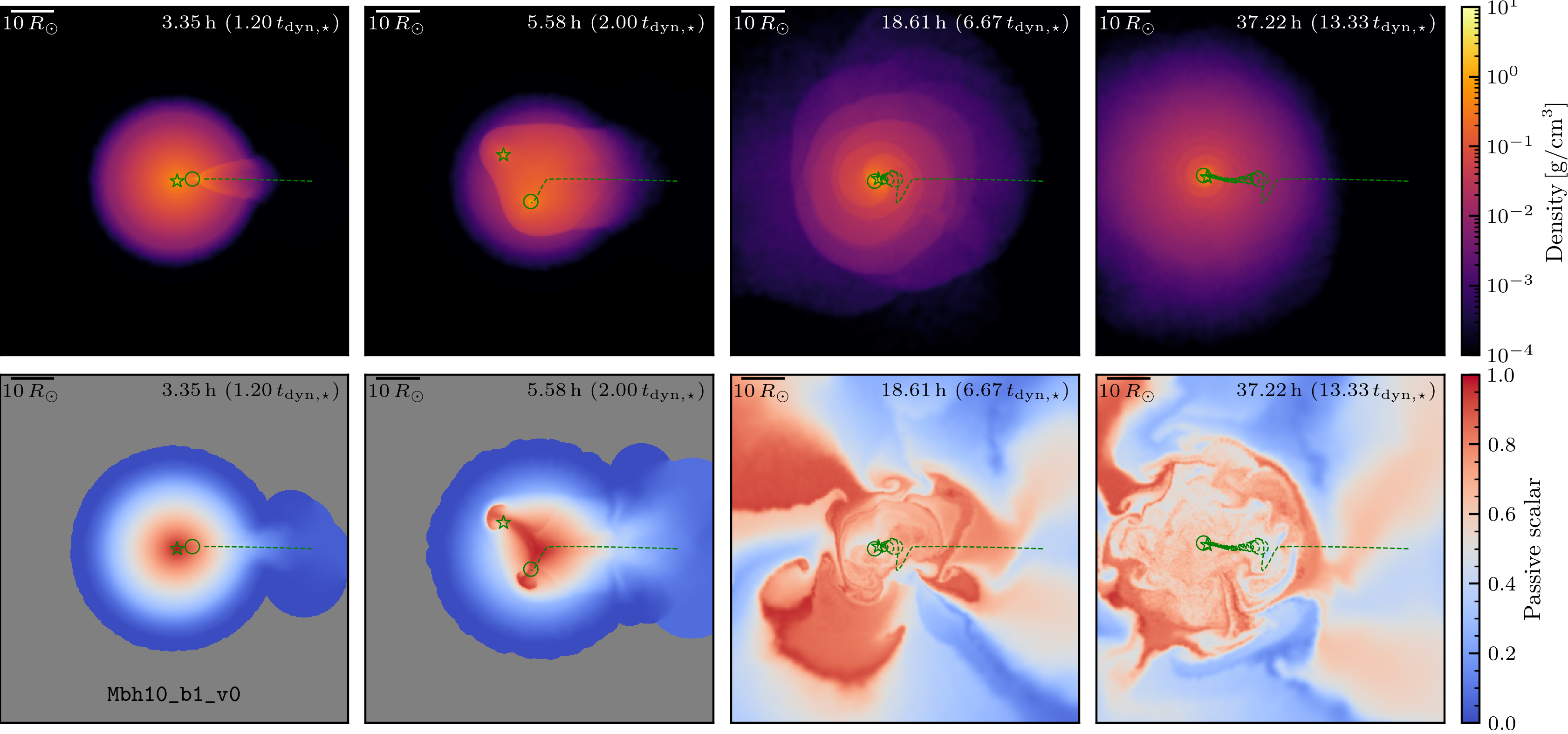}
    \caption{Visualization of a representative simulation, with central and secondary BH masses of $10\,M_\odot$, $b=R_\odot$, and $v_\infty=0$. Shown are the mid-plane ($z=0$) distributions of density (\emph{upper panels}) and passive scalar (\emph{lower panels}) at four characteristic epochs, together with the central (\emph{asterisk}) and secondary (\emph{circle}) BHs. All panels are in the lab frame. The trajectory of the secondary BH is overplotted. The passive scalar is a post-processed quantity tracking the mixing of stellar material.}
    \label{fig:vis}
\end{figure*}

\subsection{Hydrodynamical simulations}
\label{sec:method:hydro}

The hydrodynamical simulations are conducted with the Lagrangian, meshless, Godunov code \texttt{GIZMO} \citep[][]{Hopkins_2015MNRAS.450...53H} in its meshless-finite-mass (MFM) mode, with the self-gravity treated with a tree algorithm \citep[][]{HopkinsNadlerGrudic_2023MNRAS.525.5951H}. The code is configured to solve standard hydrodynamical equations with $(\rho, \bm{v}, e)$ as primitive variables (where $e$ is the internal energy) and assumes a polytropic equation of state (EOS) $p = (\gamma-1)\rho e$. 
The results presented here well complement an independent set of simulations based on the AthenaK code carried out by Sizheng Ma (in preparation).

We construct the BH* model from scratch rather than using the remnants from simulations in \citetalias{ShiHuXu_2026arXiv260827596S} (corresponding to stages~1 and 2 in Figure~\ref{fig:scenario}), mainly to ensure exact $M_{\rm \star}=100\,M_\odot$. The BH* construction begins with a $M_\star=100\,M_\odot$ and $R_\star=2\cdot 100^{1/2}\,R_\odot$ (i.e., inflated) polytrope; then we insert a central BH of $m_1$. This unstable system then undergoes a relaxation run \citep[][]{OhlmannRopkePakmor_2017A&A...599A...5O,ShiHuangLin_2026arXiv260800242S} with the \texttt{GIZMO} code, during which we add an additional damping force $\bm{a}_{\rm damp} = -\bm{v}/\tau_{\rm damp}$, where $\tau_{\rm damp} = 0.2\,(R_\star^3/GM_\star)^{1/2}$. We find that $n=2$ polytropes relax to a stable state with radius $R\approx R_\star$, while $n=3$ polytropes tend to be ultra-compact with $R\ll R_\star$. Given that our BH* formation simulations found remnants with roughly comparable central density to the progenitor star \citepalias[see][]{ShiHuXu_2026arXiv260827596S}, we choose the relaxed $n=2$ ($\gamma=1.5$) polytropic BH*.

{
In principle, our initial condition should reproduce a ``loaded polytrope'' \citep[][]{HuntleySaslaw_1975ApJ...199..328H}, i.e., a stellar envelope that is gravitated by itself and a central point mass. This is verified in Appendix~\ref{app:loaded_polytrope}.
}

We set up a suite of simulations varying $(m_1,m_2)$ and $(b,v_\infty)$. Since BH*s form only when $m_1\lesssim 0.2\,M_\star$ \citepalias{ShiHuXu_2026arXiv260827596S}, we assume $m_1, m_2 \in \{5,10\}\,M_\odot$. For each mass combination, we iterate over $b \in \{1,3,9\}\,R_\odot$ and $v_{\infty} \in \{0, 1\}\, v_0$. This ends up with 24 simulations.

For each simulation, the gaseous envelope is made of $128^3$ equal-mass resolving elements, leading to a mass resolution of $\delta m \sim 5\times 10^{-5}\,M_\odot$. This corresponds to a spatial resolution of $\delta r \sim 2R_\star/128\sim 0.2\,R_\odot$. This sets a lower limit on the BH--gas gravitational softening. As in \citetalias{ShiHuXu_2026arXiv260827596S}, we set $h_{\rm BH-gas}=0.5\,R_\odot$, while the Bondi radius is $R_{\rm Bondi} \sim G m_2/c_{\rm s}^2 \sim R_\odot \gtrsim h_{\rm BH-gas}$. BH mutual distances can be well below $R_\odot$, so for BH--BH gravitational interactions, we set $h=0$ so their gravity is accurately Newtonian in principle. {In Appendix~\ref{app:resolution_tests}, we perform a resolution convergence test with lower gas resolution ($64^3$).
}

All simulations run for $100\,(R_\star^3/GM_\star)^{1/2}\approx 140\,\rm h$.

\subsection{Semianalytic few-body simulations}
We consider a simplified limit with both BHs satisfying $m_{1,2} \ll M_\star$, and the massive star treated as a static background gas cloud. Three sources of force govern each BH's equation of motion (EOM): (1)~the gravitational force due to a static, $100\, M_\odot$ background star; (2)~the drag force when the BH is moving inside the gaseous envelope; (3)~the gravitational force due to the companion BH. The treatments of the first two sources are inherited from \citetalias{ShiHuXu_2026arXiv260827596S} and detailed there. In summary, the  $100\, M_\odot$ background star is modeled as an $n=2.75$ (dynamically stable yet $\simeq 3$) polytrope with a radius of $R_{\rm \star} = (M_\star/M_\odot)^{0.6}\,R_\odot$, and the gas drag is formulated following \citet{Ostriker_1999ApJ...513..252O} and \citet{GenerozovPerets_2023MNRAS.522.1763G}.

The third source, BH mutual interaction, is modeled by a post-Newtonian expansion up to the 2.5PN order. In particular, the 2.5PN term accounts for GW radiation that is dissipative and can shrink/circularize the orbit. For two BHs with masses $m_{1,2}$ and a total mass of $M\equiv m_1+m_2$, the GW-radiation-reaction acceleration of the relative orbit is \citep[e.g.,][]{blanchet1995gravitational}
\begin{equation}
\boldsymbol{a}_{\rm GW} = \frac{8}{5}\eta \frac{(GM)^2}{c^5 r^3} \left[ \left(3v^2 + \frac{17GM}{3r}\right)\frac{\dot{r}\boldsymbol{r}}{r} - \left(v^2 + \frac{3GM}{r}\right) \boldsymbol{v} \right],
\end{equation}
where $\eta=m_1m_2/M^2$ is the symmetric mass ratio, $\boldsymbol{r}\equiv \boldsymbol{r}_1 - \boldsymbol{r}_2$ is the relative position, and $\boldsymbol{v}\equiv\mathrm{d}\boldsymbol{r}/\mathrm{d}t$ is the relative velocity.
Therefore, the GW-radiation-reaction acceleration on $m_1$ and $m_2$ are $m_2\boldsymbol{a}_{\rm GW}/M$ and $-m_1\boldsymbol{a}_{\rm GW}/M$, respectively.

The BH orbits are solved with the $N$-body code \texttt{REBOUND} \citep{rein2012rebound} using its extension \texttt{REBOUNDx} \citep{tamayo2020reboundx}.
The EOMs are integrated using the 15th-order integrator \texttt{IAS15} \citep{ReinSpiegel_2015MNRAS.446.1424R}, and an exit minimum distance of $10^{-4}~\rsun$ is set to prevent numerical integration error.

\subsection{Caveats}
\label{sec:method:caveats}

For both hydrodynamical and semianalytic simulations, we only account for the BHs' gravity while neglecting their accretion and feedback. This assumption is based on the fact that the simulated timescale ($\sim 100$ dynamical times, $\sim 140\,\rm h$) is much shorter than the typical accretion time $\tau_{\rm Sal}\sim 0.1 \sigma_{\rm es} c/(4\pi G m_{\rm p}) \sim 45\,\rm Myr$, or the typical Kelvin-Helmholtz time $\tau_{\rm KH} \sim G M_\star^2/(2R_\star L_\star) \sim 10^4\,\rm yr$. 
\red{Besides, the time over which the BH accretes sufficient mass to liberate an amount of energy that equals the binding energy of the star is $\tau_{\rm bind} \sim (GM_{\star}^2/R_\star)/(L_{\rm Edd,\odot}M_\bullet/ M_\odot) \sim 6\times10^4$ yr, assuming Eddington-limited accretion and $M_\bullet=10\msun$.
This timescale is also much longer than the simulated timescale.}
Previous simulations \citep[][]{Murguia-BerthierMacLeodRamirez-Ruiz_2017ApJ...845..173M} also found that BH feedback is unable to disrupt stellar envelopes (common envelopes) given their high density and low compressibility. 

\red{Hydrodynamical simulations construct the BH* as a relaxed $n=2$ polytrope, while semianalytic simulations model the stellar envelope as an $n=2.75$ polytrope. However, a realistic massive-star density profile is presumably closer to that of an $n=3$ polytrope, which has a density profile that is more centrally concentrated. In Appendix~\ref{app:differentn}, we perform additional semianalytic simulations under $n=2,2.9$ and $3$ to show that the details of the density distribution does not affect our qualitative results.}

The hydrodynamical simulations do not rely on a static gaseous background and capture the evolution of both gas and BHs self-consistently. However, the main limitations are their spatial resolution and the gravitational softening for BH--gas interactions. Gas dynamics below the BH--gas gravitational softening radius are therefore unresolved. This shortcoming is complemented by models in the semianalytic simulations, allowing us to study the late-stage BBH coalescence.

\section{Result: hydrodynamical simulations}
\label{sec:hydro}

\begin{figure*}
    \centering
    \includegraphics[width=\linewidth]{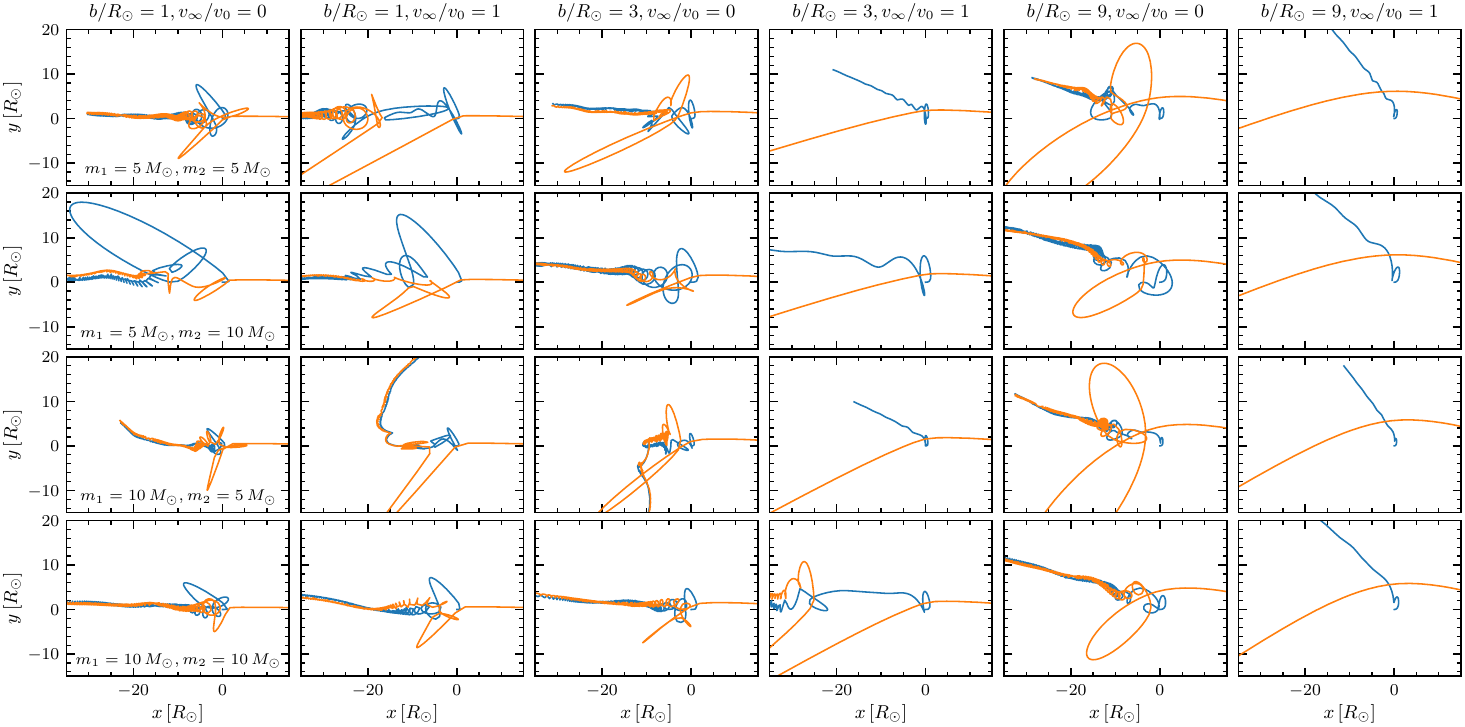}
    \caption{Trajectories of the central (\emph{blue}) and secondary (\emph{orange}) BHs for all hydrodynamical simulations performed, grouped by mass combination $(m_1, m_2)$ (rows) and labeled by impact parameter $b$ and initial velocity $v_\infty$ (columns). Cases in which a bound BH binary forms are distinguished from those in which the secondary BH escapes (see text).}
    \label{fig:hydro_trajectory}
\end{figure*}

\begin{figure*}
    \centering
    \includegraphics[width=\linewidth]{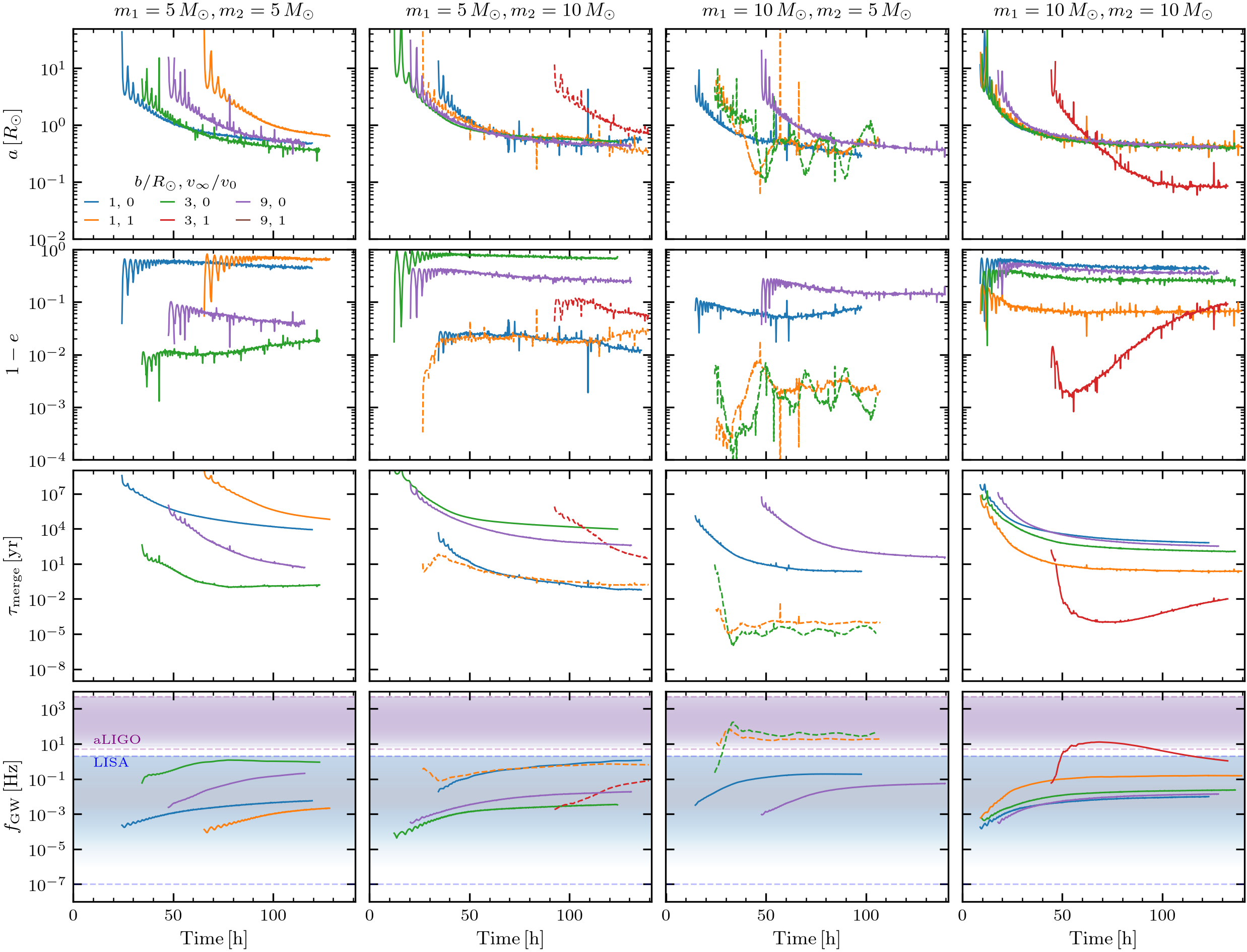}
    \caption{Time evolution of orbital semi-major axis $a$  (\emph{top row}), eccentricity (displayed as $1-e$; \emph{the second row}), GW merger time $\tau_{\rm merge}$ (\emph{the third row}), and peak GW frequency $f_{\rm GW}$ (\emph{bottom row}) for all BH binaries that form in our hydrodynamical simulations. Panels are grouped by mass combination $(m_1, m_2)$, with different curves corresponding to different initial conditions $(b/R_\odot,\, v_\infty/v_0)$ as labeled. In the bottom row, the sensitive frequency bands of aLIGO (\emph{purple}) and LISA (\emph{blue}) are overplotted, with shading proportional to $\log(1/\sqrt{S_{\rm n}})$ such that more opaque regions indicate greater detector sensitivity. { We note that a subset of simulations does not reach resolution convergence from a comparison with their counterpart low-resolution runs (see Appendix~\ref{app:resolution_tests}), so these simulations are displayed with dashed lines and for reference only.}
    }
    \label{fig:hydro_ecc}
\end{figure*}

\subsection{Dynamical formation of the BH binary}

Figure~\ref{fig:vis} visualizes a representative hydrodynamical simulation, with $m_1=m_2=10\,M_\odot$, $b=R_\odot$ and $v_\infty=0$. We show the mid-plane density and passive scalar distributions, where the passive scalar is initialized at the first snapshot as $p=(R_{\rm BH\star}-r)/R_{\rm BH\star}$, where $r$ is the distance from the central BH, and is subsequently advected with the gas without modification. The passive scalar distribution shows progressive mixing of inner stellar material toward the surface throughout the collision and subsequent BBH hardening, qualitatively consistent with the star--sBH collisions in \citetalias{ShiHuXu_2026arXiv260827596S}. 

The system undergoes four characteristic dynamical stages, corresponding to the four columns of Figure~\ref{fig:vis}. (1)~The secondary BH enters the BH*, driving a shock that simultaneously decelerates it and directs it toward the center of the potential well. (2)~By $t\approx 5.6\,\rm h$, the secondary BH passes in proximity to the central BH. Their mutual gravitational interaction ejects the central BH toward the upper-left and deflects the secondary BH toward the lower-left; both motions significantly distort the stellar envelope, as traced by the passive scalar distribution. (3)~By $t=18.6\,\rm h$, gravity draws both BHs back toward the center, where they form a bound binary. The elliptical, periodic motion of the binary drives outward-propagating shock fronts that manifest as spiral structures in the density map, while the passive scalar distribution reveals that inner material is being transported to the outer layers. (4)~By $t=37.2\,\rm h$, the binary orbit has shrunk further, the spiral arms have wound into a smoother density distribution, and the mixing of inner material toward the surface has intensified. Throughout all four stages, both BHs remain close to the center of the stellar envelope; despite the strong gravitational kicks exchanged during the close encounter, the envelope re-establishes hydrostatic balance around the binary, ensuring the BHs are continually embedded in high-density gas that sustains the drag-driven hardening.

Figure~\ref{fig:hydro_trajectory} shows the trajectories of the central and secondary BHs across all hydrodynamical simulations, spanning all possible combinations of $m_1$, $m_2$, $b$, and $v_\infty$. Not all simulations result in binary formation. Two distinct failure modes are identified. (1)~When the impact parameter is sufficiently large, the secondary BH traverses a path through lower-density regions of the BH*, reducing the integrated gas drag it experiences. If the drag is insufficient to bring the secondary BH below the escape velocity of the system, it exits the BH* without being captured. This accounts for the absence of binary formation in the $b=9\,R_\odot$, $v_\infty/v_0=1$ simulations. (2)~A light secondary BH may escape the BH*, since it experiences substantially weaker drag for the same trajectory. This is evidenced by the $b=3\,R_\odot$, $v_\infty/v_0=1$ simulations: a secondary BH of $m_2=10\,M_\odot$ is captured and forms a binary, whereas one of $m_2=5\,M_\odot$ is not sufficiently decelerated and escapes the BH*.

In all other cases, the secondary BH is retained and a bound BH binary forms at the center of the BH*. The morphology of the trajectories varies with the impact parameter: small-$b$ encounters produce sharp gravitational deflections and close interactions between the two BHs, while large-$b$ encounters yield smoother, more gradual inspiral paths. Nevertheless, binary formation occurs rapidly in both regimes.


\subsection{BH binary orbital evolution}

We study the orbital evolution of each BH binary by computing its kinetic energy $E_{\rm k} = \mu|\bm{v}_1 - \bm{v}_2|^2/2$ and potential energy $E_{\rm p} = -Gm_1m_2/|\bm{x}_1-\bm{x}_2|$, where $\mu = m_1m_2/(m_1+m_2)$ is the reduced mass. The semi-major axis then follows from the total orbital energy as $a = -Gm_1m_2/(E_{\rm k}+E_{\rm p})$, and the eccentricity from the specific angular momentum $l = |(\bm{x}_1-\bm{x}_2)\times(\bm{v}_1-\bm{v}_2)|$ via $l = [G(m_1+m_2)\,a(1-e^2)]^{1/2}$. These quantities are meaningful only once a bound binary has formed.

The top panels of Figure~\ref{fig:hydro_ecc} show the time evolution of the BH separation and semi-major axis. In simulations where no binary forms, the separation grows monotonically after the initial encounter. Where a binary does form, the separation oscillates but trends downward, and the semi-major axis decreases steadily before saturating at $a \sim 0.1$--$1\,R_\odot$ in most cases. This coincides with the softening radius for BH--gas interactions, which was caveated in Section~\ref{sec:method:caveats}.

Similarly, the second row of Figure~\ref{fig:hydro_ecc} shows the evolution of $1-e$ for all binaries that form. The majority attain high eccentricities shortly after formation, with $e \gtrsim 0.5$. In several simulations, the eccentricity is extreme, reaching $1-e \sim 0.01$--$0.1$. The most striking cases are the $m_1=10\,M_\odot$, $m_2=5\,M_\odot$ simulations with $(b/R_\odot, v_\infty/v_0) = (1,1)$ and $(3,0)$, where $1-e$ falls to $\sim 10^{-4}$--$10^{-3}$. However, we do not find a clear systematic dependence of the final eccentricity on the initial parameters $(b, v_\infty)$.

\subsection{Merger timescale and GW frequency}
\label{sec:hydro:gw}

Using the orbital parameters $(a, e)$ measured at the end of each simulation, we estimate the GW-driven merger timescale $\tau_{\rm merge}$ following \citet{Peters_1964PhRv..136.1224P}.
As shown in Figure~\ref{fig:hydro_ecc}, most binaries have $\tau_{\rm merge} \sim 1$--$1000\,\rm yr$. The extremely eccentric binaries ($1-e \sim 10^{-4}$--$10^{-3}$), however, have merger timescales orders of magnitude shorter, reaching $\tau_{\rm merge} \sim \rm seconds$--$\rm minutes$.  These binaries would therefore merge essentially instantaneously on astrophysical timescales following their formation inside the BH*.

We also compute the peak GW frequency following \citet{Wen_2003ApJ...598..419W}, who showed that an eccentric binary emits GW power across multiple harmonics of the orbital frequency, with the peak harmonic at
\begin{align}\label{eq:gw_freq}
    f_{\rm GW} = \frac{1}{\pi(1+e)^{0.3046}}\sqrt{\frac{G(m_1+m_2)}{r_{\rm p}^3}}.
\end{align}
As shown in the bottom row of Figure~\ref{fig:hydro_ecc}, most binaries have $f_{\rm GW}$ in the range $10^{-3}$--$10^{-1}\,\rm Hz$ at formation, placing them not merely within the LISA band but near its sensitivity sweet spot at $\sim 10^{-2}$--$10^{-1}\,\rm Hz$ \citep[e.g.,][]{Amaro-SeoaneAudleyBabak_2017arXiv170200786A}. Crucially, these binaries retain high eccentricities ($e \gtrsim 0.5$) throughout their evolution. Eccentric binaries are distinguishable from circularized ones through their multi-harmonic frequency structure \citep[e.g.,][]{Wen_2003ApJ...598..419W,MikocziKocsisForgacs_2012PhRvD..86j4027M,GargTiwariDerdzinski_2024MNRAS.528.4176G,chen2026PhRvD}, making them interesting targets for LISA that carry dynamical information about their formation inside a BH*.

The extremely eccentric systems ($1-e \sim 10^{-4}$--$10^{-3}$) are a qualitatively distinct population. Like the moderately eccentric binaries, they are born near the boundary between the LISA and LIGO bands; however, their very small pericenter distances drive $f_{\rm GW}$ to rapidly increase, and they subsequently enter the aLIGO/Virgo band \citep[e.g.,][]{LIGOScientificCollaborationAasiAbbott_2015CQGra..32g4001L} while still retaining extreme eccentricities. Such extreme eccentricity, once observed by LIGO, could be a distinctive signature of the BBH formation channel associated with BH*s.

\subsection{Summary and caveats}

From the hydrodynamical simulations, we find that BHs of typical masses ``capable'' of forming a BH* (in our cases, $0.05\,M_\star$ and $0.1\,M_\star$) can also be retained in a BH* if a sequential collision happens. These BH--BH*  collisions may lead to the dynamical formation of BBHs. These BBHs, once formed, can shrink their orbits over time due to the gas drag force. More distinctively, these binaries have high nascent eccentricities at formation, and some extreme cases have $e\gtrsim 0.99$. These high-eccentricity features are observable in the LISA (in extreme cases, LIGO) band even at the dynamical formation stage. 

We also perform the whole set of simulations with a lower gas resolution ($64^3$ gas cells, $\delta m\approx 4\times 10^{-4}\,M_\odot$), and find quantitatively similar results.

One important caveat is that, due to the simulation's spatial resolution, dynamics below the BH-gas softening radius ($0.5\,R_\odot$) are not resolved. With gas drag resolved at small scales, the BBH may continue to inspiral rather than exhibit the saturation seen in our hydrodynamical simulations. Additionally, the GW merging time estimated in Section~\ref{sec:hydro:gw} neglects any further gas drag effects, while the BBHs simulated here may not be GW-radiation-dominated. These limitations are complemented with the semianalytic simulations performed in Section~\ref{sec:semianalytic}.

\section{Result: semianalytic few-body simulations}
\label{sec:semianalytic}
Since the BBH formation has already been well resolved by hydrodynamical simulations, our semianalytic simulations focus on the BBH orbital evolution under the combined influence of gas drag and GW radiation.
\subsection{Case study}
We first examine a specific simulation with $m_1=5\msun$, $m_2=10\msun$, $b/\rsun=3$, $\vinf/v_0=0$ to demonstrate the physics underlying the eccentricity excitation and fast orbital decay seen in some of our hydrodynamical simulations.

\begin{figure}[htbp]
    \centering
    \includegraphics[width=\linewidth]{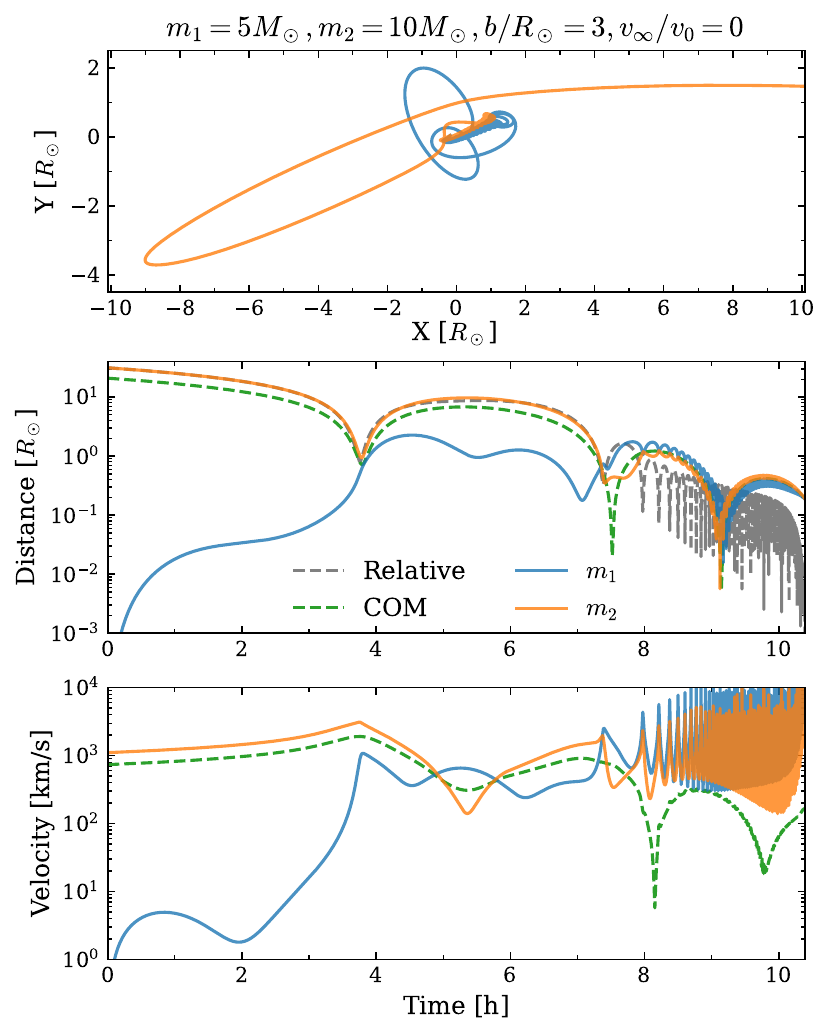}
    \caption{The trajectories (\emph{top}), positions (\emph{middle}) and velocities (\emph{bottom}) of our fiducial semianalytic simulation ($m_1=5~\msun, m_2=10~\msun, b/\rsun=3,\vinf/v_0=0$). The blue and orange lines represent the central and the intruding BH, respectively. The green dashed lines in the middle and bottom panels represent the position and velocity of the center of mass (COM) of the two BHs. The gray dashed line in the middle panel shows the relative distance between the two BHs. After $t\approx 7.3$ h, the secondary BH reaches the stellar center and becomes gravitationally bound to the central BH. 
    }
    \label{fig:traj}
\end{figure}

\begin{figure}[htbp]
    \centering
    \includegraphics[width=\linewidth]{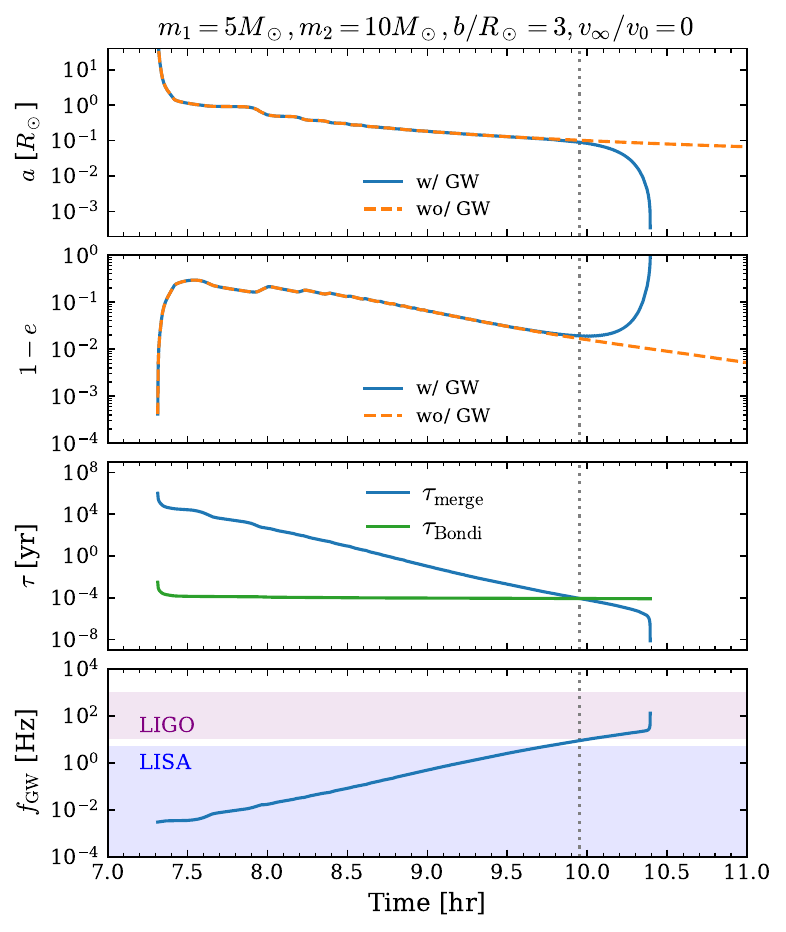}
    \caption{The semi-major axis, eccentricity, inspiral timescales, and GW frequency of the BH binary ($m_1=5~\msun, m_2=10~\msun, b/\rsun=3,\vinf/v_0=0$). The blue solid and orange dashed lines represent the semi-major axis and eccentricity of the simulations with or without the GW radiation, respectively. In the bottom row, the sensitive frequency bands of LIGO (\emph{purple}) and LISA (\emph{blue}) are overplotted. \red{The gray dotted line marks when the GW inspiral timescale equals the gas-drag inspiral timescale $\tau_{\rm merge}=\tau_{\rm Bondi}$. }
    }
    \label{fig:binary}
\end{figure}

\begin{figure*}[htbp]
    \includegraphics[width=\linewidth]{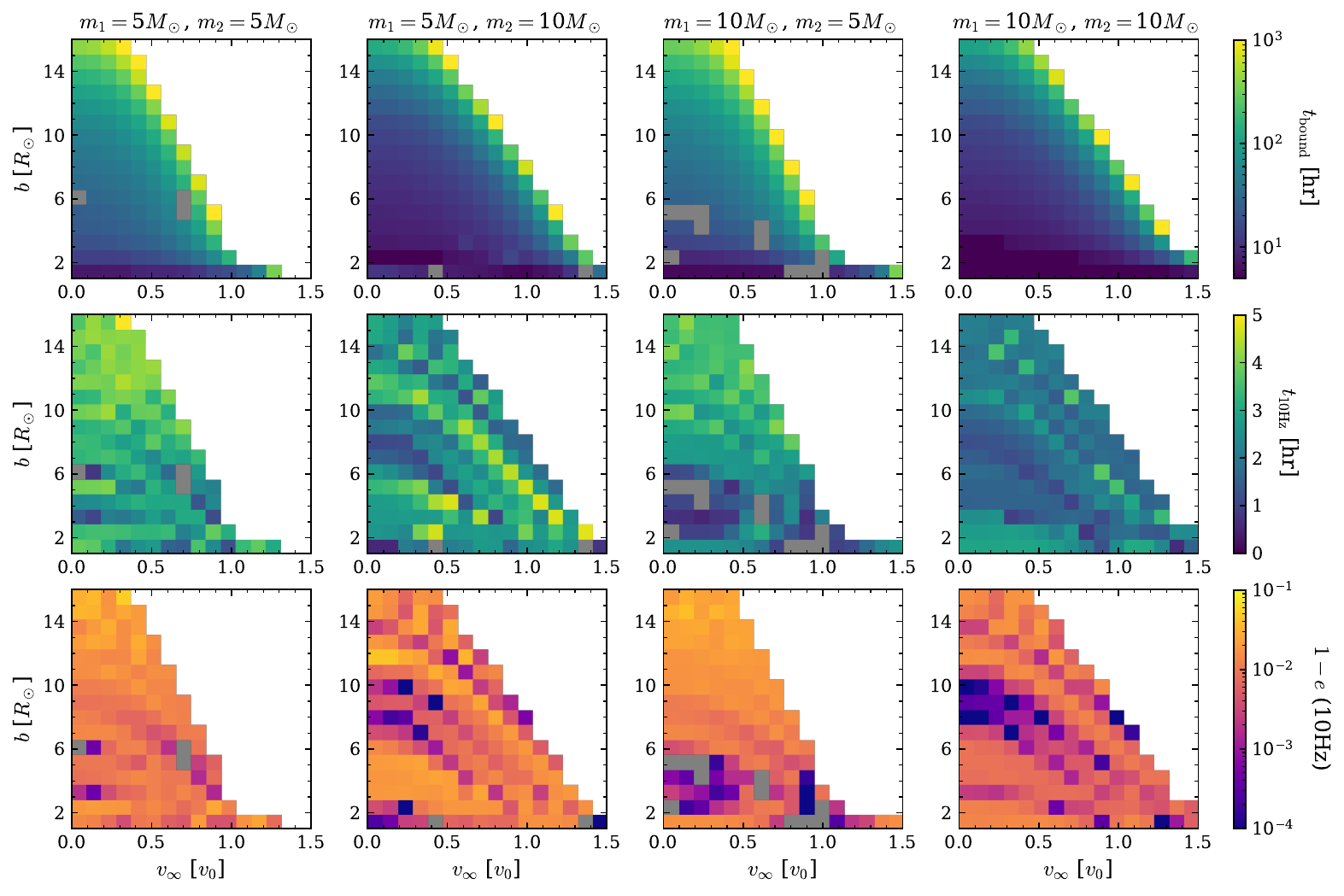}
    \caption{\emph{Top}: The time when the two BHs become bound into a binary  over the impact parameter $b$ and initial velocity $\vinf$ under different mass combinations ($m_1,m_2$). \emph{Middle}: the time elapsed between BBH formation and the GW frequency reaching 10~Hz. \emph{Bottom}: the orbital eccentricity when the GW frequency of the binary reaches 10~Hz. Blank cells show where no bound BH binary was formed within the simulation time of $10^4$ h. The gray cells indicate that the two BHs get too close to each other ($<10^{-4}~\rsun$) before they form a binary. See the relevant text for more details. 
    }
    \label{fig:grid}
\end{figure*}

\subsubsection{Gas-drag-dominated regime: binary formation and eccentricity excitation}
\label{sec:semianalytic:gas-drag}

Figure~\ref{fig:traj} shows the trajectories (\emph{top}), positions (\emph{middle}) and velocities (\emph{bottom}) of the two BHs in the fiducial system. 
At $t\approx7.3$ h, the secondary BH reaches the stellar center and becomes gravitationally bound to the central BH. The binary formation timescale here is shorter than the tens of hours from the hydrodynamical simulations, which is expected because the semianalytic simulations do not include the distortions of the stellar envelope (e.g., shocks, expansion; see Figure~\ref{fig:vis}) that could weaken the gas drag. Nevertheless, the semianalytic simulations capture the essential gas-drag dynamics and confirm the rapid BBH formation as in hydrodynamical simulations.

After the two BHs become gravitationally bound, the binary orbit shrinks continuously owing to gas drag and GW radiation.
In the first and second rows of Figure~\ref{fig:binary}, we present the evolution of the semi-major axis and the eccentricity, respectively. The blue solid line accounts for both gas drag and GW radiation, whereas the orange dashed line includes only gas drag.
At large separations (before $t\approx 10~\mathrm{h}$), the GW radiation is too weak, and hence the blue and orange lines coincide. Closer inspection reveals that before $t\approx7.5~\mathrm{h}$, both the instantaneous semi-major axis and eccentricity decrease sharply; this phase corresponds to the initial formation of the BH binary before the orbit has settled into a quasi-steady state.
Between $t\approx7.5~\mathrm{h}$ and $t\approx10~\mathrm{h}$, the semi-major axis continues to decrease steadily, while the eccentricity grows toward unity.

This eccentricity growth is ultimately driven by the characteristic behavior of gas drag, which is maximized at Mach number $\mathcal{M}\sim 1$ \citep[][]{Ostriker_1999ApJ...513..252O,KimKim_2007ApJ...665..432K,SuzuguchiSugimuraHosokawa_2024ApJ...966....7S} but drops rapidly with increasing $\mathcal{M}$ in the supersonic regime. As shown in the bottom panel of Figure~\ref{fig:traj}, the orbital velocity at periapsis is highly supersonic, whereas at apoapsis it is transonic, given the sound speed at the stellar center $\cs\approx 1000\ {\rm km\,s^{-1}}$. The force-driven eccentricity evolution for a binary orbit is given by \citep[][]{Burns_1976AmJPh..44..944B,O'NeillD'OrazioSamsing_2024ApJ...974..216O}
\begin{align}
    \frac{\dd e}{\dd t} = \frac{\sqrt{1-e^2}}{a \Omega} \left[\frac{F_r}{m} \sin f + \frac{F_\theta}{m} (\cos f + \cos E) \right],
    \label{equ:dedt_gas}
\end{align}
where $\Omega = (G M/a^3)^{1/2}$, $m$ is the orbiting mass,
$f$ is the true anomaly,  $E$ is the eccentric anomaly defined through $E-e \sin E = \Omega t$, $F_r$ and $F_\theta$ are the radial and tangential components of the driving force, respectively. At the periapsis, $\cos f + \cos E = 2$, so $\dot e \propto -2F_{\rm drag}(\mathcal{M}_{\rm peri})$; while at the apoapsis, $\cos f + \cos E = -2$, so $\dot e \propto 2F_{\rm drag}(\mathcal{M}_{\rm apo})$. For our binary, $F_{\rm drag}(\mathcal{M}_{\rm apo})>F_{\rm drag}(\mathcal{M}_{\rm peri})$ since $\mathcal{M}_{\rm apo} \sim  1$ and $\mathcal{M}_{\rm peri}\gg 1$. This naturally drives a net eccentricity growth after one period, not only because $|\dot e_{\rm apo}|>|\dot e_{\rm peri}|$, but also because the binary stays longer near the apoapsis than the periapsis. Thereby, the eccentricity is pumped to $e\to 1$ \citep[also see][]{SzolgyenMacLeodLoeb_2022MNRAS.513.5465S,O'NeillD'OrazioSamsing_2024ApJ...974..216O}, yet this effect 
will be challenged by the GW radiation, which we discuss below.

\subsubsection{GW-radiation-dominated regime: fast merger and high residual eccentricity}\label{sect:gw}

As shown in Figure~\ref{fig:binary}, the GW radiation takes over the gas drag after $t\approx 10$ h. For the orange lines without GW radiation, the semi-major axis halts at $a\sim 0.1~\rsun$, and the eccentricity continues to grow towards unity. For the blue lines with GW radiation, the semi-major axis suddenly shrinks rapidly, and the eccentricity starts to decrease due to the circularization effect of GW radiation. Within an hour after that, the orbit is circularized and the binary BHs merge.

The transition from gas-drag-dominated to GW-radiation-dominated evolution in our semianalytic simulations could be estimated by comparing the GW inspiral timescale $\tau_{\rm merge}$ following \citet{Peters_1964PhRv..136.1224P} and the inspiral timescale of the binary caused by the gas drag \citep[Equation~48 of][]{antoni2019evolution}
\begin{equation}\label{equ:tau_bondi}
    \tau_{\rm Bondi} = \frac{\eta}{\sqrt{2}} \left[ 1+\mathcal{M}_{\rm orb}^{-2} \right]^{1/2} \frac{\cs^3}{4\pi G^2 M \rho},
\end{equation}
where $\mathcal{M}_{\rm orb}\equiv v_{\rm orb}/2\cs$ is the barycentric orbital Mach number, 
$\cs/\rho$ is the local gas sound speed/density, and $\eta\approx5$ is an overall normalization factor. The $\tau_{\rm merge}$ and $\tau_{\rm Bondi}$ for the binary evolution with GW are plotted in the third row of Figure~\ref{fig:binary}. The gray dotted line marks where the two timescales are equal, which is consistent with the time when the simulations with/without GW radiation start to deviate in the $a$ and $1-e$ evolution (first and second rows of Figure~\ref{fig:binary}).

For binary evolution with GW radiation, we also plot the GW frequency (Equation~\ref{eq:gw_freq}) evolution in the fourth row of Figure~\ref{fig:binary}.
Although GW radiation starts to circularize the orbit, the binary enters the LIGO band with very high residual eccentricity, $e_{\rm 10\,Hz} \gtrsim 0.9$, well above the eccentricities inferred for observed events \citep[e.g.,][]{Romero-ShawLaskyThrane_2021ApJ...921L..31R,Romero-ShawLaskyThrane_2022ApJ...940..171R}.

\subsection{Ensemble study}\label{subsec:ensemble}

We further explore different parameters of the second BH to study whether rapid binary formation and subsequent eccentricity excitation are ubiquitous.

For the four BH mass combinations, $(m_1,m_2)=(5,5)$, (5,10), (10,5), and $(10,10)~\msun$, we run a grid of simulations with different impact parameters $b/\rsun=1,2,\cdots,16$ and initial velocities $\vinf/v_0=0,0.1,\cdots,1.5$. The results are summarized in Figure~\ref{fig:grid}.
The top panels show the color-coded time when the two BHs become bound into a binary ($t_{\rm bound}$). A blank cell indicates that no bound BH binary was formed within the simulation time of $10^4$ h. The trend of $t_{\rm bound}$ with respect to $b$ and $\vinf$ is clear: a larger $b$ or $\vinf$ leads to a longer $t_{\rm bound}$, and almost all binaries form within $\sim10^3$~h ($\sim300$ dynamical times). The gray-shaded cells indicate cases in which the two BHs approach within $<10^{-4}~\rsun$ before forming a binary. Such close encounters cannot be reliably resolved by the numerical integrator and may produce spurious ejections unless a minimum-distance termination criterion is imposed.

In the middle panels of Figure~\ref{fig:grid}, we plot the time ($t_{\rm 10\,Hz}$) elapsed between BBH formation and the GW frequency reaching 10~Hz. We find that $t_{\rm 10\,Hz} \lesssim 5\,\rm h$ for most binaries, indicating very fast coalescence after the BBH formation.

The bottom panels of Figure~\ref{fig:grid} display the orbital eccentricity $e$ at the instant when $f_{\rm GW}=10\,\rm Hz$. As is evident from Figure~\ref{fig:grid}, most binary systems exhibit $1-e_{\rm 10\,Hz} \lesssim 0.1$ (i.e., $e_{\rm 10\,Hz} \gtrsim 0.9$). This suggests a high residual eccentricity in the LIGO band for the BBHs formed through BH--BH* collisions.


\begin{figure}
    \centering
    \includegraphics[width=1.0\linewidth]{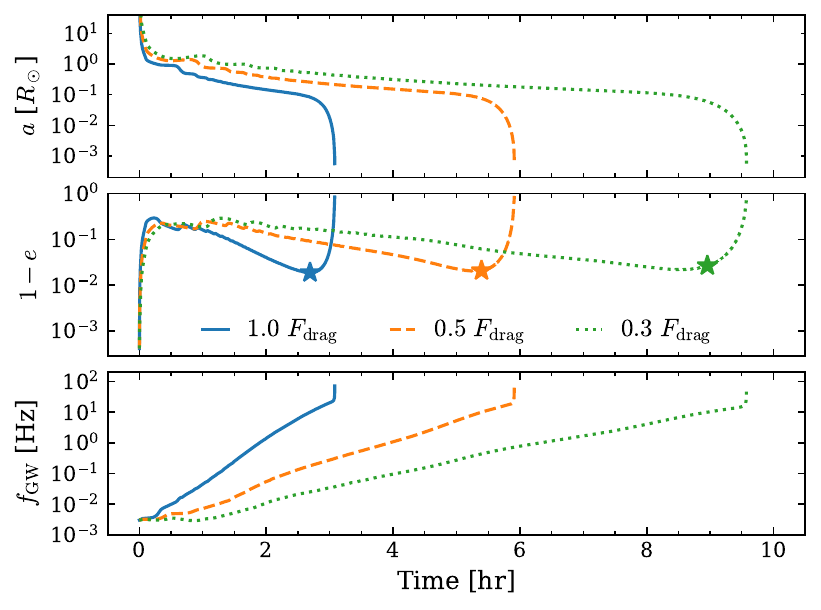}
    \caption{The BBH orbital evolution of our fiducial simulation ($m_1=5~\msun, m_2=10~\msun, b/\rsun=3,\vinf/v_0=0$) under different drag-force strengths, namely $1$, $0.5$, and $0.3$ times the fiducial drag-force magnitude \citep[][]{Ostriker_1999ApJ...513..252O}. The eccentricities at $f_{\rm GW}=10\,\rm Hz$ are marked by stars in the middle panel. The time origin is set at the moment of BBH formation. The blue solid curve corresponds to the same evolution as shown in Figure~\ref{fig:binary}.}
    \label{fig:weakdrag}
\end{figure}

\subsection{Caveats}
\label{sec:semianalytic:caveats}

Our semianalytic simulations rest on a gas-drag prescription based on the prescription of \citet{Ostriker_1999ApJ...513..252O}, which is rooted in linear perturbation theory and is expected to break down when the BH binary becomes tight and supersonic. \citet{O'NeillD'OrazioSamsing_2024ApJ...974..216O} introduced a nonlinearity parameter, $\mathcal{A}\equiv GM/(a\cs^2)$, which quantifies the ratio of the Bondi radius of a BH of mass $m$ to the binary semi-major axis $a$. In the linear regime, one requires $\mathcal{A}\ll 1$. Our binaries are clearly in the nonlinear regime, since $a\sim 1~\rsun$ at BBH formation, which is comparable to the Bondi radii of the individual BHs.

This is also related to the uncertainty about the transition from gas-drag-dominated to GW-radiation-dominated evolution. In our semianalytic simulations (Section~\ref{sect:gw}), we estimate the gas-drag inspiral timescale using the prescription of \citet{antoni2019evolution}, which was derived for binary evolution in a supersonic wind tunnel. Although our setup assumes a static background, whereas \citet{antoni2019evolution} considers a continuously flowing medium, both configurations share the feature that ample gas is available around the binary to sustain strong drag. 

In realistic stellar environments, however, the drag force may not only be weakened by nonlinear effects as the orbit shrinks, but may also, in extreme cases, be effectively suppressed if the gas is cleared out by the binary's orbital motion \citep[][]{LaiMunoz_2023ARA&A..61..517L}. If the transition to the GW-dominated regime occurs early, the gas-drag eccentricity pumping will cease prematurely, and GW circularization will lead to reduced residual eccentricity in the LIGO band.

\begin{figure*}
    \centering
    \includegraphics[width=\linewidth]{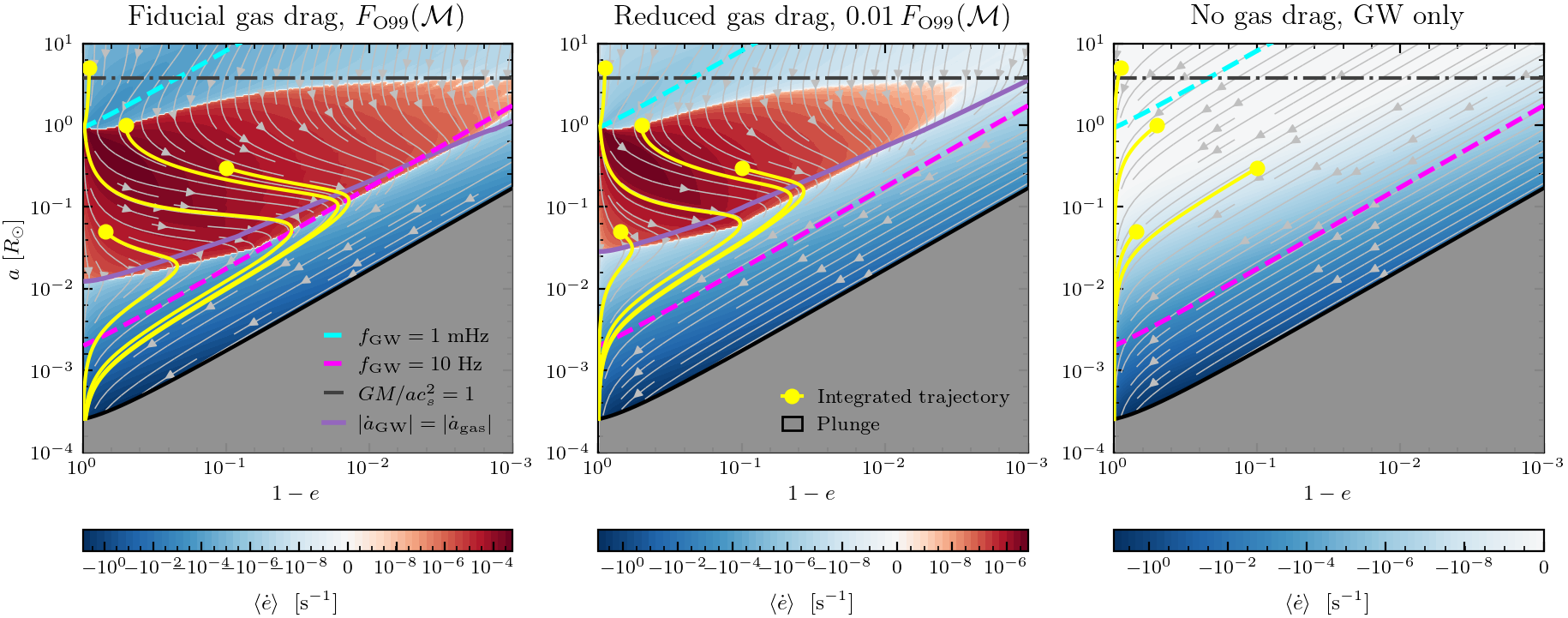}
    \caption{Phase flow in the $(a,e)$ phase space. The three panels correspond to different gas-drag magnitudes, which determine the orbital evolution along with the GW radiation \citep[][]{Peters_1964PhRv..136.1224P}. Each panel shows the phase flow and contours of $\langle \dot e\rangle$. We also pick 4 typical initial conditions (\emph{yellow dots}) and show their evolution in the phase space (\emph{yellow lines}). The gray shaded region is inaccessible, set by the Schwarzschild separatrix $r_{\rm p}\equiv a(1-e) = (6+2e)GM/c^2$ \citep[][]{CutlerKennefickPoisson_1994PhRvD..50.3816C}. We also show the contours of $f_{\rm GW}$ at $1\,\rm mHz$ (typical LISA band; \emph{cyan dashed line}) and $10\,\rm Hz$ (typical LIGO band; \emph{magenta dashed line}). 
    }
    \label{fig:ae_phase_space}
\end{figure*}

\section{Discussion}
\label{sec:discussion}

\red{Here we first discuss the gas-drag eccentricity pumping, then some missing pieces in the full scenario (cf. Figure~\ref{fig:scenario}), and finally the possibility of subsolar-mass mergers inside a less massive star.}

\subsection{Gas-drag eccentricity pumping}

As discussed in Section~\ref{sec:semianalytic:caveats}, BBH orbital evolution in the nonlinear regime remains uncertain. We defer a detailed exploration of the transition point from gas-drag-dominated to GW-radiation-dominated evolution to future work. However, to assess the impact of weaker gas drag on the BBH orbital evolution, we evolve the binary in our fiducial setup under different drag-force normalizations, namely $1$, $0.5$, and $0.3$ times the fiducial drag magnitude \citep{Ostriker_1999ApJ...513..252O}, \emph{after} the BBH formation. Figure~\ref{fig:weakdrag} shows the corresponding evolution of the semi-major axis, eccentricity, and GW frequency. While reducing the drag force delays the transition to the GW-radiation-dominated regime, the residual eccentricity at $10~\mathrm{Hz}$ remains close to unity ($e_{\rm 10\,Hz}>0.9$) for all cases considered. This indicates that the qualitative outcome, high residual eccentricity, is robust against these drag force magnitudes.

We further study the gas-drag eccentricity pumping in the phase space of $(a,e)$. Under the drag force, the eccentricity evolves following Equation~\eqref{equ:dedt_gas}, while the semi-major axis follows \citep[][]{Burns_1976AmJPh..44..944B,O'NeillD'OrazioSamsing_2024ApJ...974..216O}
\begin{align}
    \frac{1}{a}\frac{\dd a}{\dd t} = \frac{2}{a \Omega \sqrt{1-e^2}} \left[\frac{F_r}{m}e \sin f + \frac{F_\theta}{m} (1 + e\cos f) \right].
    \label{equ:dadt_gas}
\end{align}
These equations apply at specific phases of an orbit, and the orbit-averaged evolution rate is then defined by, e.g., $\langle \dot e\rangle = T^{-1} \int_0^T \dot e \dd t$. Combining Equations~\eqref{equ:dedt_gas}, \eqref{equ:dadt_gas}, and the orbit-averaged evolution ($\langle \dot a\rangle_{\rm GW}$, $\langle \dot e\rangle_{\rm GW}$) given by \citet{Peters_1964PhRv..136.1224P}, we obtain the net orbit-averaged evolution: $\langle \dot a \rangle = \langle \dot a\rangle_{\rm gas} + \langle \dot a\rangle_{\rm GW} $ and $\langle \dot e \rangle = \langle \dot e\rangle_{\rm gas} + \langle \dot e\rangle_{\rm GW} $. In real calculations, we note that each BH feels its own drag force, so $F_{r}$ and $F_\theta$ in Equations~~\eqref{equ:dedt_gas} and \eqref{equ:dadt_gas} need to account for the difference in the drag forces.

Figure~\ref{fig:ae_phase_space} presents $\langle\dot e\rangle$ for three different assumptions on the gas drag force magnitude: fiducial \citep[][]{Ostriker_1999ApJ...513..252O}, 0.01 times the fiducial, and none (i.e., GW-only). The figure assumes typical numbers at the stellar core, $\rho=1\,\rm g\,cm^{-3}$ and $c_{\rm s}=1000\,\rm km\,s^{-1}$. For each panel, we also overplot the phase flow in the $(a,e)$ plane that is given by $\langle \dot a \rangle$ and $\langle \dot e \rangle$. Without gas drag, $\langle \dot e \rangle<0$ across the plane, so the phase flow monotonically converges to $e=0$. However, since gas drag can excite the eccentricity growth (as qualitatively inferred in Section~\ref{sec:semianalytic:gas-drag}), there is an ``island'' of $\langle\dot e\rangle>0$ in the $(a,e)$ plane. The upper boundary of the island is close to $\mathcal{A} = G M/(a c_{\rm s}^2)=1$, i.e., when the BBH enters the nonlinear regime in the gaseous medium; it is also the boundary where the BBH motion becomes transonic. The lower boundary, however, is close to the transition from gas-drag dominance to GW-radiation dominance, where $\dot a_{\rm GW}=\dot a_{\rm gas}$. Beyond this transition, the orbits shrink but circularize. Inside the eccentricity-pumping island, the eccentricity of the BBH increases monotonically.

Comparing the two cases with drag force, we find that with our fiducial prescription \citep[][]{Ostriker_1999ApJ...513..252O}, the transition from gas-drag dominance to GW-radiation dominance coincides with the boundary where $f_{\rm GW}=10\,\rm Hz$. That explains the high residual eccentricity of the BBHs. However, with a much weaker drag force, $0.01$ times the fiducial, the BBHs enter the GW-radiation-dominated regime early. Their residual eccentricities at $f_{\rm GW}=10\,\rm Hz$ are much lower, even though they had a high natal eccentricity or were once pumped to high eccentricity due to gas drag. 

As a result, we find that the high residual eccentricity generally requires the BBH orbit to remain gas-drag-dominated when it enter the LIGO band. This holds when the drag force is still at the order of classical \citet{Ostriker_1999ApJ...513..252O} magnitude (or slightly smaller, $0.3\times$, cf. Figure~\ref{fig:weakdrag}), but is no longer valid if the drag force is significantly weaker (e.g., $0.01\times$). Still, the LISA band ($\sim 1\,\rm mHz$) remains in the regime where the gas-drag eccentricity pumping is effective: LISA can therefore possibly witness the eccentricity excitation of these BBHs if the event is bright enough.

\subsection{Retention of the merged BH}
Whether the merged BH remains gravitationally bound to the massive star or is ejected is determined by the competition between the GW recoil velocity ($v_{\rm kick}$) imparted to the remnant BH and the escape velocity ($v_{\rm esc}$) of the host star. For a massive star with mass $M_\star=100~\msun$ and radius $R_\star=16\,\rsun$, the surface escape velocity is approximately $\approx 1500~{\rm km}\,{\rm s}^{-1}$, while the escape velocity from the center is $v_{\rm esc} = \sqrt{-2\Phi(r=0)} \approx 2800\,\rm km\,s^{-1}$. The magnitude of the recoil kick depends sensitively on the masses and spins of the progenitor BHs. Particularly for BH*s, the central BH is accreting at a highly convective core \citep[][]{BegelmanRossiArmitage_2008MNRAS.387.1649B,CoughlinBegelman_2024ApJ...970..158C,HassanPernaCantiello_2026ApJ...998...65H}. As demonstrated by \citet{chen2023chaotic}, BHs embedded in such a turbulent environment undergo chaotic gas accretion: the orientation of the accretion flow is randomized with respect to the BH spin axis on an eddy-turnover timescale. Consequently, any spin angular momentum acquired by the BH through prior mergers or coherent accretion episodes is efficiently suppressed by the stochastic torques exerted by the turbulent inflow. This mechanism drives the central BH toward low spins before it encounters the incoming BH.

\red{Given the low central-BH spin expected from chaotic accretion, aligned spins with
$\chi_{1z},\chi_{2z}\lesssim0.1$ produce
$v_{\rm kick}\lesssim200~{\rm km}\,{\rm s}^{-1}$
\citep{islam2026accurate}. 
However, if the second BH has a significant in-plane spin component, NR simulations can produce, for example, $v_{\rm kick}\sim1800~{\rm km}\,{\rm s}^{-1}$ for $\chi\simeq0.8$ \citep{Boyle:2019kee}.
This exceeds the surface escape velocity $\approx1500~{\rm km}\,{\rm s}^{-1}$, but remains below the central escape velocity, $v_{\rm esc}\approx2800~{\rm km}\,{\rm s}^{-1}$.
Since the BBH generally merges near the stellar center,} the merger remnant can still hardly escape the gravitational potential of the host star. Instead, it will undergo dynamical friction against the stellar gas and sink back toward the stellar center on a few stellar dynamical timescales \citepalias[][]{ShiHuXu_2026arXiv260827596S}. Once settled, this heavier BH becomes a new seed for subsequent hierarchical collisions, awaiting further inspiral and merger with additional BHs that may collide with the newly formed BH*.


\subsection{Massive second BH}\label{sec:massive_intruder}

The analysis presented above focuses on stellar-mass BHs with $M_\bullet\lesssim 10~\msun$, for which the dissipated orbital energy remains subdominant compared with the gravitational binding energy of the host star, so the binary inspiral proceeds inside a dense, extended gaseous envelope. However, if the second BH is significantly more massive \citepalias[$\gtrsim0.2\,M_\star$ as suggested in][]{ShiHuXu_2026arXiv260827596S}, the dissipated orbital energy can overcome the gravitational binding energy of the host star. Instead of a stable BH* state within a dynamical time, the envelope undergoes large-scale mass ejection and may appear as a micro-TDE \citep[e.g.][]{PeretsLiLombardi_2016ApJ...823..113P,KremerLombardiLu_2022ApJ...933..203K}.

In this regime, the embedded binary may no longer evolve inside a spherical envelope but rather within a circumbinary disk (CBD). The CBD modifies the binary's orbital evolution in several important ways. Gravitational torques exerted by the non-axisymmetric density distribution in the CBD can extract angular momentum from the binary \citep{escala05,siwek23,ChenLin_2024ApJ...967...88C}, while gas accretion onto the individual BHs and dynamical friction against the disk further alter the semi-major axis and eccentricity \citep{tagawa20,lichenlin2024,rowan25, pan2026, ida2026}. Depending on the binary mass ratio and orbital eccentricity, CBD torques can either accelerate the inspiral toward merger or, for low-mass-ratio systems, cause temporary orbital expansion \citep{kleynelson2012, siwek23}. Consequently, the merger timescale and the final spin-orbit geometry of a BH binary assembled in this way may differ significantly from the vacuum or spherical-envelope cases discussed in Sections~\ref{sec:hydro} and \ref{sec:semianalytic}.

Repeated mergers increase the mass of embedded black holes.  
The capture of the primary black holes with $M_\bullet \gtrsim 0.2 
M_\star$ leads to the dispersal of the target stars \citepalias[][]{ShiHuXu_2026arXiv260827596S}.
In an AGN disk (where the density is many orders of magnitude
smaller than that of the stellar envelope), gas drag effect on
embedded black holes is much weaker. The binary-capture 
probability of bare black holes \citep{lilai2022, lidempsey2023}
is much lower than that induced by main-sequence-star conduits.  
Loosely bound binary black holes may be hardened by
their host AGN disks \citep{baruteau2011, lidong2022}
on timescales much longer than the dynamical timescales inside
main-sequence stars.  For typical masses (up to $\sim 10^3 M_\odot$) 
of either metamorphic \citep{Ali-DibLin_2023MNRAS.526.5824A, XuChenLin_2026ApJ...997..206X} or immortal stars \citep{CantielloJermynLin_2021ApJ...910...94C}, the growth
of BHs' mass through mergers in embedded main-sequence
stars in AGN disks may be quenched with a drop-off in the 
mass function above $M_\bullet \sim 10^2 M_\odot$.

\subsection{The formation rate of BBHs through BH--BH* collisions in AGN disks}
{
Preliminary analysis of stellar population 
in AGN disks, including its dynamical evolution,
have been carried out by
\citet{ChenLin_2024ApJ...967...88C}.  A series
of follow-up analyses and simulations 
are contained in papers
already submitted or in preparation elsewhere.  
We briefly recapitulate here the basic physics of AGN 
stars and justify the formation rate of BBHs through BH--BH* collisions.

In typical AGN disks, self-regulated star formation
generates sufficient flux $Q^+ _\star$ to balance
radiative flux from the disk surface $Q^-$. In the 
outer ($\gtrsim 0.1$pc) regions of the disk, embedded
stars attain an accretion-wind equilibrium
with a top-heavy mass function.  With insufficient
mixing across their radiative layers \citep{xu2025fate}, 
the embedded stars evolve off the main sequence on a
timescale $\tau_{\rm ms} \sim$ a few Myr, contaminate 
the disk with $\alpha$ elements and Fe-peak byproducts, 
and produce multiple ($\sim \tau_{\rm AGN}/\tau_{\rm ms}$) 
generations of stellar-mass BH remnants \citep{xu2025stellar} 
during the active AGN episode $\tau_{\rm AGN} \sim 10^8$ yr.  
These metamorphic stars' average mass, luminosity, and 
radius are $M_\star \sim$ a few $10^2 M_\odot$, $L_\star 
\sim L_{\rm Edd, \odot} M_\star/M_\odot$, and 
$R_\star \sim$ a few $10 R_\odot$, where $M_\odot$, 
$R_\odot$, and $L_{\rm Edd, \odot}$ 
are the mass, radius, and Eddington-limited luminosity 
of the Sun. 

In the radiation-pressure-dominated,
optically-thick regions of the disk, 
\begin{equation}
    Q^- \simeq {4 c c_{\rm s} \Omega \over \kappa} \simeq {L_{\rm Edd, SMBH} h \over \pi R_{\rm disk}^2}
    \label{eq:qminus}
\end{equation}
where $L_{\rm Edd, SMBH} = L_{\rm Edd,
\odot} M_{\rm SMBH}/M_\odot$ and $M_{\rm SMBH}$ are the SMBH's Eddington-limited luminosity and mass, $\kappa$ is the 
electron-scattering opacity, 
$\Omega= (G M_{\rm SMBH}/R_{\rm disk}
^3)^{1/2}$, $c_s$ and $h$ are the mid-plane sound speed 
and aspect ratio at disk radius $R_{\rm disk}$. 
A thermal equilibrium ($Q^+ _\star = Q^-$)
is maintained with the surface number density of stars satisfying
\begin{equation}
s_\star \simeq h M_{\rm SMBH}  /\pi R_{\rm disk}^2 M_\star.
\end{equation}
Multiple generations of star formation produce
$s_\star < s_\bullet \lesssim s_\star \tau_{\rm AGN}/\tau_{\rm ms}$. 

Assuming that a similar velocity dispersion $\sigma_\star \sim c_s$ 
is established for the stars, BHs, and BH*s through
their mutual interaction and drag by the disk gas
\citep{ChenLin_2024ApJ...967...88C}, we estimate the
collision frequency of a star with other stars to be 
\begin{equation}
    \Gamma_{\rm \star} 
    (R_{\rm disk}) \simeq 
    \pi (1+ \Theta) s_{\rm \star} R_\star^2 \Omega,
\end{equation}
where $\Theta= G M_\star / R_\star \sigma_\star^2 \sim 
{\mathcal O} (10^2)$ is the Safronov number
\citep{BinneyTremaine_2008gady.book.....B}.
This estimate is consistent with that obtained from
N-body simulations (Wang et al. in preparation).

On a stellar collision timescale $\tau_{\star} \simeq 
\Gamma_\star ^{-1}$, internal mixing leads to the replenishment
of H-laden gas into the stellar cores to prolong their
main sequence evolution (Shi et al. submitted).  
Within $R_{\rm disk} \lesssim 10^{3-4} R_\bullet$ 
(where $R_\bullet= G M_{\rm SMBH}/c^2$
is the SMBH's gravitational radius), $\tau_{\star} 
\leq \tau_{\rm ms}$ \citep{ChenLin_2024ApJ...967...88C}
and episodic stellar collisions indefinitely
prolong their main sequence life span (Xu et al. 
in preparation).  The average $M_\star$, $L_\star$, 
and $R_\star$ of these immortal stars 
\citep{CantielloJermynLin_2021ApJ...910...94C}
are larger than those of metamorphic stars by at most a factor of two \citep{Ali-DibLin_2023MNRAS.526.5824A}.
Although immortal stars do not produce and yield 
additional $\alpha$ elements, H burning through the
CNO cycle can enhance the N/(C+O) abundance ratio 
\citep{xu2025stellar}.  The observed 
metallicity gradient in AGN disks (Huang et al. 
submitted) provides supporting evidence 
for the coexistence of such immortal stars in the inner 
disk with metamorphic stars in the outer disk (Cao et 
al. submitted).  

Although immortal stars do not evolve off the main 
sequence, BHs produced by the metamorphic stars
in the outer disk diffuse into the inner regions of 
the disk through stochastic migration \citep{wu2024}.
The collision frequency of a star with a BH is 
\begin{equation}
    \Gamma_{\bullet} \simeq 
    \pi (1+ \Theta) s_{\bullet} R_\star^2 \Omega 
    \sim \Gamma_{\star}.
\end{equation}
At this rate of BH* formation, a significant 
fraction of embedded stars may have captured
black hole remnants such that $s_{\rm BH*} \sim s_\star$.

Under the assumption that the structure of BH* is
not significantly affected by the engulfment of a 
much lower-mass ($q < 0.2$) BH core \citep[][Xu et al
in preparation]{ShiHuXu_2026arXiv260827596S}, we expect the BH--BH* collision 
frequency $\Gamma_{\rm BH*} \sim \Gamma_{\bullet} 
\sim \Gamma_\star$.
Since the velocity dispersion is small ($\sigma_\star\ll v_0$), encounters are nearly parabolic ($\vinf/v_0\sim\sigma_\star/v_0\to 0$). Consequently, even modest energy dissipation can render the incoming BH's orbital energy negative, allowing BH--BH* collisions to efficiently capture a second low-mass BH ($q<0.2$) across the range of impact parameters considered.
Our order-of-magnitude estimate on the formation rate of 
BBHs through BH--BH* collision over the entire AGN disk  
\begin{equation}
    \Gamma_{\rm BBH*} \simeq 2 \pi \int
    s_{\rm BH*} \Gamma_{\rm BH*} (R_{\rm disk})
    R_{\rm disk} d R_{\rm disk}
\end{equation}
suggests it is a significant fraction of the LIGO event rate.
But we have not yet taken into account the possible
outcomes of collisions between embedded stars and 
more massive ($q > 0.2$) BHs (see discussions in Section~\ref{sec:massive_intruder}). More comprehensive 
population analyses are beyond the scope of this paper 
and their comparison 
with the LIGO data will be presented elsewhere.
}

\subsection{Subsolar-mass mergers inside a less massive star}
\red{
Subsolar-mass compact objects have been proposed to arise from exotic formation channels, including primordial black holes
\citep{hawking1971,carr1974}, subsolar-mass neutron stars \citep[e.g.,][]{metzger2024ApJ}, etc. Such subsolar-mass compact binaries have also been targeted by dedicated gravitational-wave searches \citep[e.g.,][]{2023MNRAS.524.5984L}.}

\red{An intriguing possibility is that a lower-mass star ($\sim 1\,\msun$) could capture such subsolar-mass compact objects \citep[e.g.,][]{GottliebCantielloNorton_2026arXiv260602700G,CantielloGottliebNorton_2026arXiv260602726C}, with strong gas drag within its envelope subsequently driving their inspiral and merger. This would extend the scenario explored in this work for $\sim 100\,\msun$ stars and $\sim 10\,\msun$ black holes to lower-mass systems.
Exploring this possibility would require assessing the availability of such compact objects and their interactions with lower-mass stellar envelopes, whose structures may differ from those considered here.
}

\section{Conclusions}
\label{sec:conclusions}

In this work, we have investigated a dynamical formation channel for highly eccentric compact binaries in which sBHs are sequentially captured by massive main-sequence stars. This is based on the finding in \citetalias{ShiHuXu_2026arXiv260827596S} that low-mass sBHs ($M_\bullet\lesssim 0.2\,M_\star$) can be efficiently captured by massive stars through gas dynamical friction, settling into quasi-hydrostatic BH*s that preserve most of the original stellar mass. Using a combination of three-dimensional hydrodynamical simulations and semianalytic few-body simulations, we have characterized the formation, orbital evolution, and GW signatures of black-hole binaries assembled inside the gaseous stellar envelopes. Our main conclusions are as follows.

\begin{enumerate}
    \item \textbf{High natal eccentricity.} The impulsive nature of the BH--BH* collision imprints a high eccentricity on the nascent binary. Our hydrodynamical simulations show that binaries are readily born with $e\gtrsim 0.5$, and its orbital frequency is already within the LISA sensitivity band. In extreme cases, the eccentricity reaches $1-e\sim 10^{-4}{-}10^{-3}$ at formation. 

    \item \textbf{Gas-drag-driven eccentricity pumping.} After formation, the binary orbit shrinks under the domination of gas dynamical friction. With the BBH orbiting at $\mathcal{M}\sim 1$ near the apoapsis and $\mathcal{M}\gg 1$ near the periapsis, gas drag preferentially drives more eccentricity growth near the apoapsis than eccentricity damping near the periapsis \citep[][]{Burns_1976AmJPh..44..944B,Ostriker_1999ApJ...513..252O,KimKim_2007ApJ...665..432K}, thereby monotonically pumping the eccentricity toward unity \citep[][]{SzolgyenMacLeodLoeb_2022MNRAS.513.5465S,O'NeillD'OrazioSamsing_2024ApJ...974..216O}. 

    \item \textbf{High residual eccentricity in the LIGO band.} GW radiation then dominates over the gas drag at the late stage, causing the tendency to circularize the BBH orbit. However, our semianalytic models demonstrate that the residual eccentricity remains high ($e_{10\,\mathrm{Hz}}>0.9$) in the LIGO sensitivity band, before GW radiation eventually circularizes the orbit during the final coalescence phase.

    \item \textbf{Rapid BBH merger.} The transition from gas-drag-dominated to GW-radiation-dominated evolution occurs at separations $a\sim 0.1\,R_\odot$, after which the binary merges within $\sim$hours. Most binaries in our ensemble study reach the LIGO band ($f_{\rm GW}\approx 10\,$Hz) within a few hours after formation, guaranteeing a prompt merger.

    \item \textbf{Retention of the merger remnant.} The progenitor BH can be driven to low spins by chaotic gas accretion in the turbulent stellar interior \citep{chen2023chaotic}, and the GW recoil kick remains modest ($v_{\rm kick}\lesssim 200\,\mathrm{km}\,\mathrm{s}^{-1}$; \citealt{islam2026accurate}), far below the escape velocity \red{ from the star center ($v_{\rm esc}\approx 2800\,\mathrm{km}\,\mathrm{s}^{-1}$)}. The merger product therefore remains gravitationally bound, sinks back to the stellar center via dynamical friction, and becomes a seed for subsequent hierarchical mergers.
\end{enumerate}

The impulsive collision assumed in this scenario is essential to ensure a high natal eccentricity at the BBH formation. This high natal eccentricity then triggers the gas-drag-driven pumping once the BBH moves transonically (supersonically) at the apoapsis (periapsis). The incoming BH may still be absorbed by the BH* through a nearly circular inspiral orbit, but the low natal eccentricity may not be pumped to unity by the gas drag. The impulsive collision also suggests that such BBHs can be much rarer than typical BBHs since the direct geometric collision rate is limited by the tiny cross section ($\sim R_\star^2$) compared with binary capture. 

The same mechanism may also be applicable to the T\.ZO \citep[][]{ThorneZytkow_1977ApJ...212..832T}, to form more generally defined compact binaries (NSBH, NSNS). 

Our analysis is restricted to secondary intruding BHs with $M_\bullet\lesssim 10\,\msun$ (i.e., $M_\bullet\lesssim 0.1\,M_\star$). For more massive intruders ($M_\bullet\gtrsim 0.2\,M_\star$), the dissipated orbital energy exceeds the stellar binding energy, preventing the formation of a stable BH*; the remnant is instead a dynamically unstable, disk-like structure or a micro-TDE \citepalias[][]{ShiHuXu_2026arXiv260827596S}. In such cases, binary inspiral proceeds within a circumbinary disk, requiring a separate treatment.

There are several important astrophysical implications. For example, the high residual eccentricities predicted for this channel, particularly $e_{\rm 10\,Hz}\gtrsim 0.9$ in the LIGO band, constitute a distinctive observable signature that could discriminate BH--BH* mergers from binaries formed in the field or through purely gravitational few-body scattering, even compared with existing eccentric BBH formation channels \citep[][]{Samsing_2018PhRvD..97j3014S,TagawaKocsisHaiman_2021ApJ...907L..20T,SamsingBartosD'Orazio_2022Natur.603..237S}. The retention of merger products furthermore enables repeated star--sBH collisions as a pathway for rapid BH growth in dense stellar systems, potentially contributing to intermediate-mass black-hole formation in nuclear star clusters. A detailed exploration of the transition from gas-drag-dominated to GW-radiation-dominated evolution, including nonlinear hydrodynamical effects at small separations, is deferred to future work.

\begin{acknowledgments}
We thank the anonymous referee for comments that helped improve this article.
We thank Sizheng Ma and Hongxuan Jiang for sharing the results of their simulations and close interaction.  We also benefited from useful conversations with Long Wang, Huan Yang, and Yixian Chen.
YS and NM acknowledge the support of the Natural Sciences and Engineering Research Council of Canada (NSERC) under funding reference number 568580.
QH acknowledges the support of the National Natural Science Foundation of China (grant Nos.\ 12173021 and 12133005). 
Computations were performed on the ``Trillium'' supercomputer at the SciNet HPC Consortium. SciNet is funded by Innovation, Science and Economic Development Canada; the Digital Research Alliance of Canada; the Ontario Research Fund: Research Excellence; and the University of Toronto.
\end{acknowledgments}

%

\vspace{5mm}


\software{\texttt{GIZMO} \citep[][]{Hopkins_2015MNRAS.450...53H};
        \texttt{REBOUND} \citep[][]{rein2012rebound};
        \texttt{REBOUNDx} \citep[][]{tamayo2020reboundx}
        }



\appendix

{
\section{Loaded polytropes}
\label{app:loaded_polytrope}

\begin{figure}
    \centering
    \includegraphics[width=\linewidth]{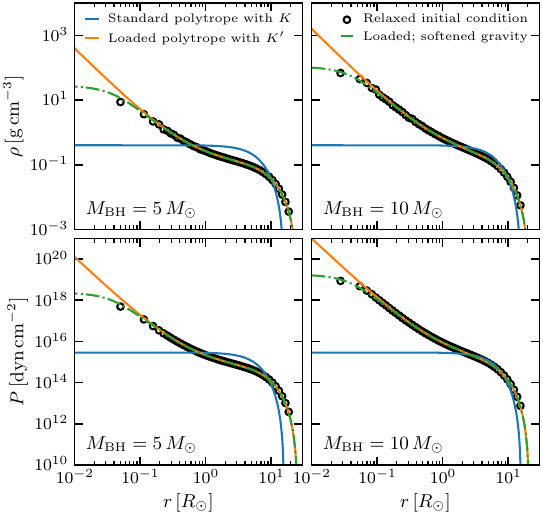}
    \caption{Density and pressure profiles of different stellar and BH* models: (1)~a standard $n=2$ polytrope (\emph{blue}); (2)~the BH* initial conditions adopted in our hydrodynamical simulations (\emph{circles}); (3)~the unsmoothed loaded polytrope (\emph{orange}); and (4)~the loaded polytrope with softened central gravity (\emph{green dot-dashed}). We use different polytropic constants $K$ for the standard and loaded polytropes (see Appendix~\ref{app:loaded_polytrope}).}
    \label{fig:polytrope_comparison}
\end{figure}

In this section, we compare the initial conditions adopted in our hydrodynamical simulations (Section~\ref{sec:method:hydro}) with loaded polytrope models \citep[][]{HuntleySaslaw_1975ApJ...199..328H}. Unlike standard polytropes \citep[][]{Lane_1870AmJS...50...57L}, loaded polytropes contain a central point mass $M_{\rm BH}$. We assume a polytropic equation of state (EOS), $P=K\rho^\gamma$, where $\gamma=1+1/n$. The specific enthalpy is then $h=\int \dd P/\rho=(n+1)K\rho^{1/n}$, giving $\rho(h)=[h/((n+1)K)]^n$. The hydrostatic structure equations are
\begin{align}
    \frac{\dd h}{\dd r} &= -\frac{G M(r)}{r^2}, \\
    \frac{\dd M}{\dd r} &= 4\pi r^2\rho(h).
\end{align}
Here, $M(r)$ is the total enclosed mass, including the central BH. The loaded polytrope satisfies the boundary conditions $h(R_\star)=0$, $M(R_\star)=M_{\rm BH}+M_\star$, and $M(0)=M_{\rm BH}$, where $R_\star$ and $M_\star$ are the stellar radius and mass, respectively. These equations can be solved numerically to obtain $h(r)$ and $M(r)$.

The density profile depends on the polytropic index $n$. Near the center, the gravitational potential of the central point mass dominates, giving $h\sim GM_{\rm BH}/r$ and hence $\rho(r)\propto r^{-n}$. Consequently, the stellar mass enclosed within radius $r$ scales as $M(r)-M_{\rm BH}\propto r^{3-n}$ ($n\neq 3$). A finite stellar mass near the center therefore requires $n<3$. However, even for $n<3$, the density diverges as $r\to 0$, producing a central cusp in the unsmoothed loaded polytrope.

In our simulations, we soften the BH's gravitational potential to avoid the numerical divergence associated with the central point mass. Assuming Plummer softening, the hydrostatic equation becomes
\begin{align}
    \frac{\dd h}{\dd r}
    = -\frac{G M(r)r}{(r^2+\epsilon^2)^{3/2}},
\end{align}
where $\epsilon$ is the gravitational softening length. In the limit $r\to 0$, the softened gravitational acceleration vanishes, so $h(r)$ approaches a finite constant and the density likewise approaches a finite central value. The softened model therefore develops a central core rather than the cusp found in the unsmoothed solution.

Figure~\ref{fig:polytrope_comparison} compares the BH* initial conditions used in our hydrodynamical simulations with the standard and loaded polytrope models, both with and without gravitational softening. We also show a standard $n=2$ polytrope with a mass of $100\,M_\odot$ and a radius of $100^{0.6}\,R_\odot$, which uniquely determines the polytropic constant $K$ \citep[see, e.g.,][]{KippenhahnWeigertWeiss_2013sse..book.....K}. As described in Section~\ref{sec:method:hydro}, our initial conditions adopt the value of $K$ corresponding to a standard polytrope with a mass of $100\,M_\odot$ and a radius of $2\times100^{0.6}\,R_\odot$. We denote this value by $K'$ in Figure~\ref{fig:polytrope_comparison} and use it to construct both the unsmoothed and smoothed loaded polytropes. In our simulations, we adopt a gravitational softening length of $\epsilon=0.04\,R_\odot$, approximately corresponding to the mean spatial resolution. The profiles show good agreement between our initial conditions and the smoothed loaded polytrope. Although the unsmoothed loaded polytrope exhibits a steep central cusp, its profile agrees with our initial conditions at $r\gtrsim0.1\,R_\odot$.

\section{Resolution convergence tests}
\label{app:resolution_tests}

\begin{figure}
    \centering
    \gridline{
        \fig{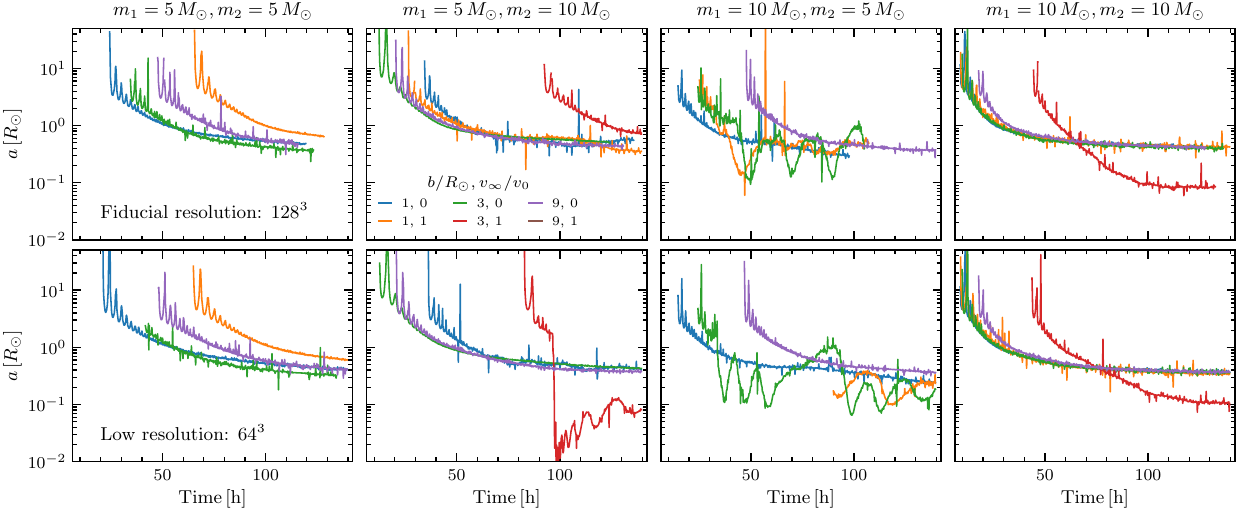}{\linewidth}{(a) Comparison of the semi-major axis.}
    }
    \vspace{-10pt}
    \gridline{
        \fig{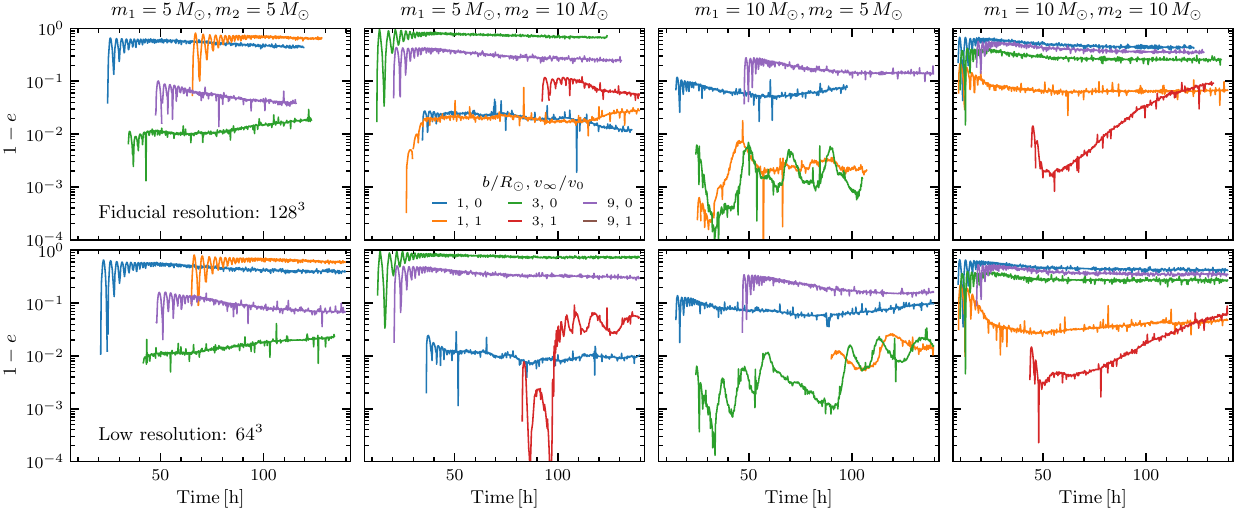}{\linewidth}{(b) Comparison of the eccentricity.}
    }
    \caption{Resolution comparison of the orbital evolution of BBHs, showing the semi-major axis (top) and eccentricity (bottom) in our fiducial ($128^3$) and lower-resolution ($64^3$) simulations. The simulations are initialized with the same combinations of $(m_1,m_2,b,v_\infty)$ at both resolutions. The comparison assesses the sensitivity of the binary orbital evolution to numerical resolution.}
    \label{fig:resolution_convergence}
\end{figure}

In this section, we investigate the resolution dependence of our simulations by comparing the fiducial ($128^3$) simulations with a suite of lower-resolution ($64^3$) runs. The two sets of simulations use the same physical setup and initial orbital parameters (see Section~\ref{sec:method:hydro}); the only difference is the resolution of the initial conditions and the corresponding hydrodynamical calculations. We consider the same combinations of $(m_1,m_2,b,v_\infty)$ as in Figures~\ref{fig:hydro_trajectory} and \ref{fig:hydro_ecc}, allowing us to compare the resulting orbital evolution directly.

Figure~\ref{fig:resolution_convergence} compares the evolution of the binary semi-major axis $a$ and eccentricity $e$ between the two resolutions. Overall, we find good agreement for most of the simulations, suggesting that the main features of the orbital evolution are reasonably robust to changes in resolution. In particular, for the cases with $m_1=m_2=10\,M_\odot$, the low- and high-resolution simulations show good agreement in both $a$ and $e$ for all five runs that form BBHs. This agreement indicates that, for these cases, the formation and subsequent orbital evolution of the binaries are not strongly affected by the change in resolution.

However, noticeable differences emerge in a subset of simulations involving $5\,M_\odot$ stellar-mass BHs (sBHs). A common feature of these runs is that the eccentricity can approach unity with $1-e\sim 10^{-3}$, such that the periapsis is as small as $\sim 10^{-3}\,R_\odot$ (given $a \sim R_\odot$ at BBH formation), which is well below the gas resolution scale. These runs should be ultimately assessed with even higher-resolution simulations. As a result, in Figure~\ref{fig:hydro_ecc} we distinguish these simulations with dashed lines. However, our main conclusion, that high-eccentricity BBHs can form due to BH--BH* collisions, is still guaranteed by other simulations that passed the resolution convergence test.

\section{Polytropes of different $n$}
\label{app:differentn}

\begin{figure*}
    \centering
    \includegraphics[width=\linewidth]{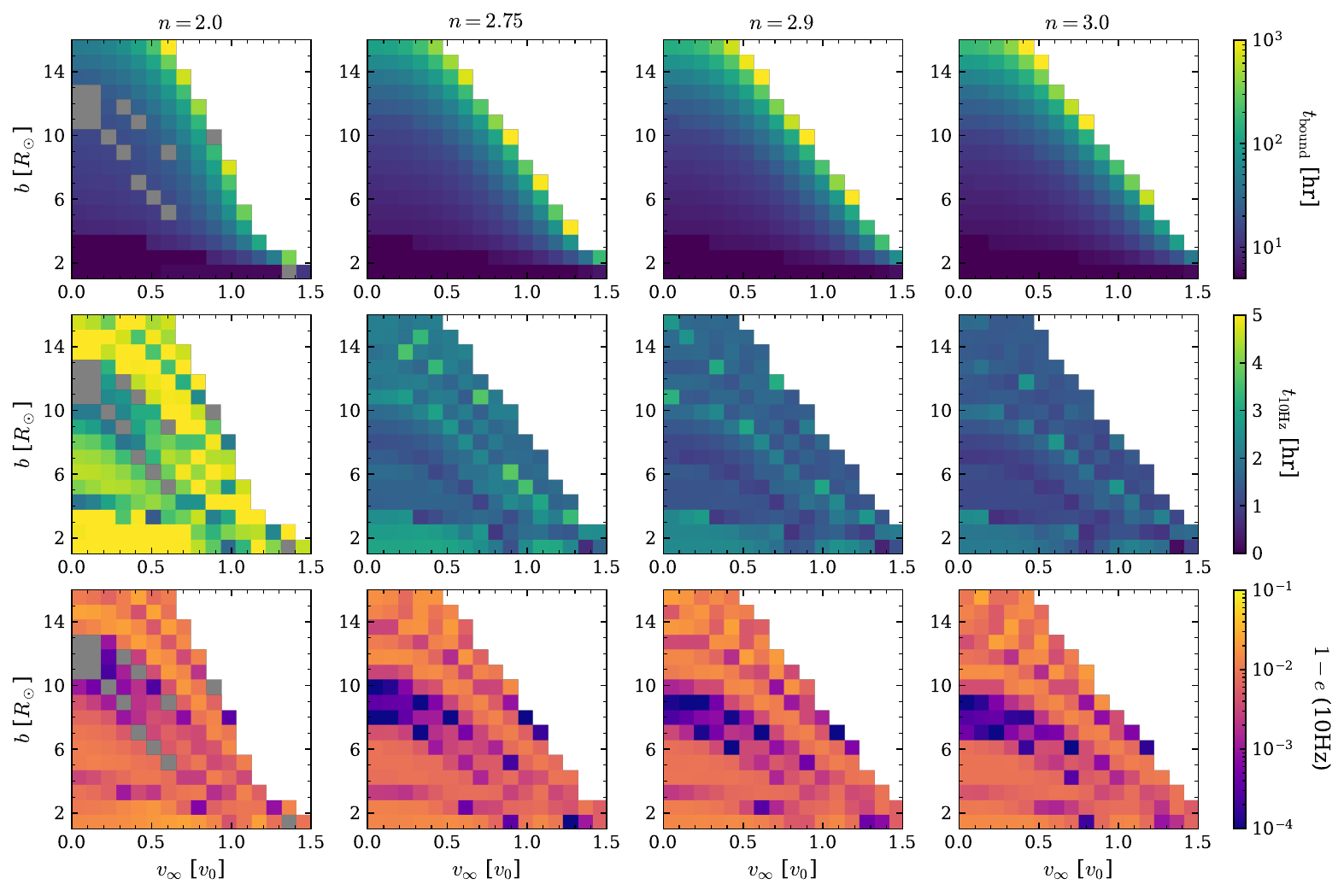}
    \caption{Similar to Figure~\ref{fig:grid}, but for different polytropic indexes ($n$) under fixed BH masses ($m_1=10\msun,m_2=10\msun$).}
    \label{fig:grid-n}
\end{figure*}

In the main text, we construct the massive star as an $n=2$ polytrope in hydrodynamical simulations and an $n=2.75$ polytrope in semianalytic simulations. However, a realistic massive-star density profile is presumably closer to that of an $n=3$ polytrope, which has a density profile that is more centrally concentrated.

In this section, we conduct complementary semianalytic ensemble studies as in Section~\ref{subsec:ensemble} for massive stars with different polytropic indexes $n$ to investigate the influence of the stellar density profile on the inspiral of the second BH. The stellar mass and radius ($M_\star=100\msun,R_\star\approx16\rsun$) are the same as in the main text, and the BH masses are fixed to $m_1=10\msun$ and $m_2=10\msun$. Figure~\ref{fig:grid-n} shows the results for $n=2.0,2.75,2.9$ and $3.0$. The $n=2.75$ column here is the same as that in Figure~\ref{fig:grid}. 
For $n=2$, there exist many gray grids in which the separation between the two BHs falls below $10^{-4}\rsun$. This occurs because the $n=2$ stellar model retains most of its mass at large radius, allowing the second BH to rapidly lose orbital angular momentum and fall nearly radially toward the first BH at the stellar center, resulting in a very close encounter.

Nevertheless, the retention boundary of the second BH, the time when two BHs become bound (top panels), and the orbital eccentricity at $10$~Hz (bottom panels) are generally the same across different polytropic indexes.
The time elapsed between BBH formation and the GW frequency reaching $10$~Hz (middle panels) significantly decreases as $n$ increases from 2 to 3, because the higher central densities of polytropes with larger $n$ produce stronger gas drag, accelerating orbital shrinkage.

}

\bibliography{bh,gw,star,hqr}{}
\bibliographystyle{aasjournal}



\end{document}